\documentclass{iopart}

\usepackage{graphicx}
\usepackage{epsfig}
\usepackage{ifthen,iopams,setstack,amssymb}
\begin{document}
\title[Exact spin dynamics of the $1/r^2$ SUSY {\itshape t-J} model in a magnetic field]{
Exact spin dynamics of the $1/r^2$ supersymmetric {\itshape t-J} model in a magnetic field}

\author{Mitsuhiro Arikawa$^1$, Yasuhiro Saiga$^2$}

\address{$^1$ Yukawa Institute for Theoretical Physics, Kyoto University, Kyoto 606-8502, Japan}
\address{$^2$ Department of Physics, Nagoya University, Nagoya 464-8602, Japan}

\begin{abstract}
The dynamical spin structure factor $S^{zz}(Q,\omega)$ in the small momentum region is derived analytically for the one-dimensional supersymmetric {\itshape t-J} model with 
$1/r^2$ interaction. Strong spin-charge separation is found in the spin dynamics.
The structure factor $S^{zz}(Q,\omega)$ with a given spin polarization does not depend on the electron density
in the small momentum region.
In the thermodynamic limit, only two spinons and one antispinon (magnon) contribute to $S^{zz}(Q,\omega)$.
These results are derived via solution of the SU(2,1) Sutherland model in the strong coupling limit.   

\end{abstract}
%Uncomment for PACS numbers title message
% Keywords required only for MST, PB, PMB, PM, JOA, JOB? 
%\vspace{2pc}
%\noindent{\it Keywords}: Article preparation, IOP journals
% Uncomment for Submitted to journal title message
\submitto{\JPA}
% Comment out if separate title page not required
%\maketitle

%  YITP-06-10 (preprint number in YITP)
%

\section{Introduction}
The spin-charge separation is a main subject
in one-dimensional interacting electron systems.
The conformal field theory has succeeded in
the description of spin-charge separation in the low-energy physics of the Tomonaga-Luttinger liquid.
Beyond the conformal field theory limit, 
exactly solvable models provide us chances
to obtain analytical knowledge on thermodynamics and dynamics,
and it is intriguing how the spin-charge separation appears in these properties.

Among exactly solvable models,
the supersymmetric $t$-$J$ model with $1/r^2$ interaction
\cite{kuramoto91} reveals the spin-charge separation in the simplest manner.
The Hamiltonian of this model is given by
\begin{eqnarray}
\hspace{-2cm}
{\mathcal H}_{tJ} & = & 
\sum_{i<j}
\left[
-t_{ij}\sum_{\sigma=\uparrow,\downarrow}
\left(\tilde{c}^{\dagger}_{i\sigma}\tilde{c}_{j\sigma}+h.c.\right)
+J_{ij}
\left(
\mbox{\boldmath $S$}_{i}\cdot\mbox{\boldmath $S$}_{j}
-\frac{1}{4}n_{i}n_{j}
\right)
\right]  - h \sum_{j} S_j^z,
\label{tj-hamiltonian}
\end{eqnarray} 
where $\tilde{c}_{i\sigma}=c_{i\sigma}(1-n_{i,-\sigma})$ with
$c_{i\sigma}$ being the annihilation operator of an electron with spin
$\sigma$ at site $i$, and
$n_{i}
=
\sum_{\sigma}n_{i,\sigma}=\sum_{\sigma}c^{\dagger}_{i\sigma}c_{i\sigma}$.
The spin operator associated with site $i$ is defined as
$\mbox{\boldmath $S$}_{i}
=\sum_{\alpha,\beta}
c^{\dagger}_{i\alpha} 
(\mbox{\boldmath $\sigma$}_{i})_{\alpha\beta}
c_{i\beta}/2$
where $\mbox{\boldmath $\sigma$}=(\sigma_x,\sigma_y,\sigma_z)$
is the vector of Pauli matrices.
The transfer energy $t_{ij}$ and exchange one $J_{ij}$ are given by
$t_{ij}=J_{ij}/2=t D^{-2}_{ij}$ where
$D_{ij}=(N/{\pi})\sin\left(\pi\left(i-j\right)/N\right)$
with $N$ being the number of lattice sites.
Henceforth we take $t$ as the unit of energy.
We note that this {\itshape t-J} model reduces to the 
Haldane-Shastry spin chain model~\cite{haldane88,shastry88} at half-filling.
In the supersymmetric {\itshape t-J} model with $1/r^2$ interaction,
exact thermodynamics
can be interpreted in terms of 
free spinons and holons \cite{kuramotokato95,katokuramoto96}.
At low temperature, the spin susceptibility is independent of
the electron density $\bar{n}$,
and the charge susceptibility is independent of
the magnetization $\bar{m}$.
These features are referred as to the strong spin-charge separation~\cite{kuramotokato95}. 
In addition,
the fact that the magnetization $\bar{m}$ for a certain range of $h$ is independent of $\bar{n}$ can be regarded as the strong spin-charge separation.
Namely, $\bar{m}$ is determined by $h$ as follows:
\begin{equation}
\bar{m} = \left\{ 
\begin{array}{ll}
1-\sqrt{1-2h/\pi^2}, & \mbox{for $0\le h \le h_c$} \\
\bar{n}, & \mbox{for $h \ge h_c$}    
\end{array}
\right.
\end{equation}
where $h_c = \bar{n}(2-\bar{n})\pi^2/2$ \cite{kawakami}.  

The strong spin-charge separation appears also in dynamics at zero temperature.
The dynamical spin structure factor is given by
\begin{equation}
\label{defii}
S^{zz}(Q,\omega)
=
\sum_{\nu}
|\langle\nu|
S^{z}_{Q}
|0 \rangle|^2
\delta(\omega-E_{\nu}+E_0),
\end{equation}
where 
$S^z_{Q}
=\sum_l S^z_{l}e^{iQl}/\sqrt{N}$.
Here $|\nu \rangle$ 
denotes a normalized eigenstate of the Hamiltonian
with energy $E_\nu$ ($E_0$ being the ground state energy).
In the absence of magnetic field ($h=0$), 
the dynamical spin structure factor was exactly obtained 
at $\bar{n}=1$~\cite{haldane93,yamamoto00a,yamamoto00b}.
It was numerically demonstrated that
the weight of the dynamical spin structure factor in the {\it t-J} model
does not depend on $\bar{n}$ in the region
where only two spinons contribute~\cite{saiga99}.
This is an indication of the strong spin-charge separation in dynamics.
Later, a mathematical poof was given to this statement,
and the analytical expression of the dynamical spin structure factor
for $\bar{n}<1$
was obtained in the full $(Q,\omega)$ space~\cite{arikawa04a}.

A numerical study~\cite{saiga00} also showed that
the strong spin-charge separation for $S^{zz}(Q,\omega)$ can be extended to the case of finite magnetic field ($h \ne 0$).
Namely, at fixed magnetization, $S^{zz}(Q,\omega)$ away from half-filling is the same as that for half-filling (i.e., the Haldane-Shastry model),
in the region where only spinons and antispinons contribute.
For $h \ne 0$,
the full exact results on $S^{zz}(Q,\omega)$ have not been obtained
even in the Haldane-Shastry model.
However, if the momentum is restricted to $Q  \le \pi \bar{m}$,
the dynamical structure factor $S^{zz}(Q,\omega)$ at $\bar{n}=1$ can be expressed as the dynamical density-density correlation function of the Sutherland model
with coupling parameter $\beta=2$ \cite{talstra94,arikawa00}.
In order to give this expression, we assume the positive magnetization $\bar{m}$ without loss of generality,
where $\bar{m}=\bar{n}_\uparrow-\bar{n}_\downarrow$ with $\bar{n}_\sigma = N_\sigma/N$
($N_\sigma$ being the number of electrons with $\sigma$-spin). 
In the thermodynamics limit, we have the following expression:
\footnote{For the expression of $S^{zz}(Q,\omega)$ in Ref.\cite{arikawa00}, 
the authors erroneously typed the integration ranges of the spinon momenta $q_i (i=1,2)$ as $0 < q_i < \pi \bar{n}_\downarrow$.
They should read $0 < q_i < \pi (1 - \bar{m})$, as shown in Eq.(\ref{f:thermodynamiclimit}).}
\begin{eqnarray}
S^{zz}(Q,\omega) & = & 
\frac{Q^2}{4\pi} \int_{0}^{\pi (1-\bar{m})} d q_1 \int_{0}^{\pi (1-\bar{m})} d q_2
\int_{0}^{\pi \bar{m}} d q_{\rm a}  \nonumber \\
& & \times 
\delta\left(Q-q_a-\sum_{j=1}^2 q_j \right) 
\delta\left(\omega-\epsilon_{\rm a}(q_a)-\sum_{j=1}^2 \epsilon_{\rm s}(q_j)
\right) \nonumber \\
& & \times \frac{|q_1-q_2| \ \ \ \epsilon_{\rm a} (q_{\rm a})}
{\prod_{j=1}^2 (q_{\rm a} + 2q_j)^2 
\prod_{j=1}^2 \epsilon_{\rm s}(q_j)^{1/2}},
\label{f:thermodynamiclimit}
\end{eqnarray}
where $\epsilon_{\rm s}(q)$ is the spinon spectrum: 
$\epsilon_{\rm s}(q) = q(v_{\rm s}-q)$, 
and $\epsilon_{\rm a}(q)$ is the antispinon spectrum:  
$\epsilon_{\rm a}(q) = q(v_{\rm s}+q/2)$,
where $v_{\rm s}=\pi (1-\bar{m})$.
The purpose of this paper is to prove that
the analytical expression of $S^{zz}(Q,\omega)$ away from half-filling is the same as Eq.\ (\ref{f:thermodynamiclimit}),
if $0 < Q \le \min[\pi \bar{m}, \pi \bar{n}_\downarrow]$.
This yields a mathematical proof of the strong spin-charge separation in magnetic-field dynamics.
We stress that the use of the replica type technique is crucial
for calculation of the matrix element in $S^{zz}(Q,\omega)$.

This paper is organized as follows.
In the next section, we introduce the SU(2,1) Sutherland model as an auxiliary. 
As in the previous study of thermodynamics~\cite{kuramotokato95,katokuramoto96,polychronakos90,sutherland93} and dynamics~\cite{yamamoto00a,yamamoto00b,arikawa99,peysson03}, we take the limit $\beta \rightarrow \infty$ of the coupling parameter 
in order to obtain the analytical knowledge of
the $1/r^2$ supersymmetric {\itshape t-J} model. 
The eigenfunctions of the Sutherland model can be expressed in terms of Jack polynomials. 
We discuss the basic features of the  Jack polynomials.  In section \ref{sec.matrixelement}, we derive the matrix element of the dynamical spin structure factor based on the replica type technique \cite{gangardt01}. In section \ref{sec.result}, we present the 
analytic expression of $S^{zz}(Q,\omega)$ for finite systems.
Section \ref{sec.summary} is devoted to summary.
In \ref{app.nqw}, we derive the dynamical charge structure factor $N(Q,\omega)$ in the same method.
In \ref{app.comparison}, we show the comparison with numerical results 
for small size systems \cite{saiga00}.
In \ref{app.static}, we present the results on the static structure factors $S^{zz}(Q)$ and $N(Q)$.   

\section{Sutherland model with SU(2,1) symmetry}
In this section, we introduce the Sutherland model 
\cite{sutherland1,sutherland2,sutherland3,sutherland4} with SU(2,1) symmetry \cite{katokuramotoSu},
and review the basic properties.
\subsection{Notation}
We follow the notations of Refs.\cite{macdonald95,kato98,arikawa04b}.  
For a fixed
non-negative integer $n$, let
$\Lambda_n
=
\{\eta=(\eta_1,\eta_2,\cdots,\eta_n)\, |
\,\eta_i\in{\bf Z}_{\ge 0},\,1\leq i\leq n\}$
be the set of all compositions with length less
than or equal to $n$.
The diagram of a composition
$\eta=(\eta_1,\eta_2,\cdots,\eta_n)\in\Lambda_n$
is defined as the set of points
$(i,j)\in\mbox{\bf{Z}}^2$ such that $1\le j\le \eta_i$.
The weight $||\eta||$ of a composition
$\eta=(\eta_1,\eta_2,\cdots,\eta_n)\in\Lambda_n$ is defined by
$||\eta||=\sum_{i=1}^n\eta_i$.
The length $l(\eta)$ of $\eta$ is defined as the
number of non-zero $\eta_i$ in $\eta$.  The set of all partitions with
length less than or equal to $n$ is defined by
$\Lambda_n^+
=\{\lambda=(\lambda_1,\lambda_2,\cdots,\lambda_n)\in\Lambda_n\,
|\,\lambda_1\geq\lambda_2\geq\cdots\geq\lambda_n\geq 0\}$.
We also denote a partition $\lambda$ by $a^{m_a}b^{m_b}c^{m_c}\cdots$
or by $(a^{m_a},b^{m_b},c^{m_c},\cdots)$ with
$a>b>c>\cdots \ge 0$
where $m_i$ is the number of parts which are equal to $i$.
The conjugate partition $\lambda'$ of a partition $\lambda$ is a
partition whose diagram is the transposition of the diagram
of $\lambda$.
Hence $\lambda_i'$ is the number of nodes in the $i$-th column of
the diagram of partition $\lambda$. In particular we have
$\lambda_1'=l(\lambda)$.
We define the subset $\Lambda_n^{+,>}$ of the set $\Lambda_n^{+}$ by
$\Lambda_n^{+,>}
=\{\lambda=(\lambda_1,\lambda_2,\cdots,\lambda_n)\in\Lambda_n\,
|\,\lambda_1>\lambda_2>\cdots>\lambda_n\geq 0\}$.
Notice that for any element $\lambda\in\Lambda_n^{+,>}$,
there exists unique partition
$\mu\in\Lambda_n^+$ such that $\lambda=\mu+\tilde{\delta}(n)$ with
$\tilde{\delta}(n)=(n-1,n-2,\cdots,1,0)\in\Lambda_n^{+,>}$.  For two
distinct partitions $\lambda$, $\mu\in \Lambda^+_n$, we define the
dominance order $\lambda<\mu $ if $\vert\vert\lambda\vert\vert=\vert\vert\mu\vert\vert$
and $\sum_{i=1}^k\lambda_i\le\sum_{i=1}^k\mu_i$ for all
$k=1,\cdots,n$.  For a composition $\eta\in\Lambda_n$, $\eta^+$ denotes
the unique partition which is a rearrangement of the composition $\eta$.
Now we define a partial order $\prec$ on compositions as follows: for
$\nu,\eta\in\Lambda_n$, $\nu \prec\eta$ if $\nu^+ <\eta^+$ with
dominance ordering on partitions or if $\nu^+=\eta^+$
and $\sum_{i=1}^k\nu_i\le\sum_{i=1}^k\eta_i$ for all $k=1,\cdots,n$.

For a given composition $\eta=(\eta_1,\eta_2,\cdots,\eta_n)$
and $s=(i,j)$ in the diagram of the composition $\eta$, we define
the following quantities:
\begin{eqnarray}
\label{arm}
a_\eta(s)
&=&
\eta_i-j,
\\
\label{co-arm}
a'_\eta(s)
&=&
j-1,
\\
\label{leg}
l_\eta(s)
&=&
\#\{k\in\{1,\cdots,i-1\}|j\leq\eta_k+1\leq\eta_i\}
\nonumber
\\
&&
+\#\{k\in\{i+1,\cdots,n\}|j\leq\eta_k\leq\eta_i\},
\\
\label{co-leg}
l'_\eta(s)
&=&
\#\{k\in\{1,\cdots,i-1\}|\eta_k\geq\eta_i\}
\nonumber
\\
&&
+\#\{k\in\{i+1,\cdots,n\}|\eta_k>\eta_i\}.
\end{eqnarray}
Here, for a set $A$, $\#A$ denotes the number of elements.  The 
quantities $a_\eta(s)$, $a'_\eta(s)$, $l_\eta(s)$ and $l'_\eta(s)$ are
called arm-length, arm-colength, leg-length, and leg-colength,
respectively. Since $l'_\eta(s)$ does not depend on $j$, we also denote
it as $l'_\eta(i)$.
Notice that for a partition $\lambda\in\Lambda_n^+$, we
have
\begin{eqnarray}
\label{leg'}
l_\lambda(s)
&=&\lambda'_{j}-i,
\\
\label{co-leg'}
l'_\lambda(s)
&=&i-1.
\end{eqnarray}

Further, for a composition $\eta\in\Lambda_n$ and real parameters $r$ and
$\gamma$, we define the following quantities:
\begin{eqnarray}
\label{c}
f_{\eta}(r;\gamma)
&=&
\prod_{s\in\eta}(a'_\eta(s)-rl'_\eta(s)+\gamma),
\\
\label{d}
d_{\eta}(r)
&=&
\prod_{s\in\eta}(a_\eta(s)+1+r(l_\eta(s)+1)),
\\
\label{d'}
d'_{\eta}(r)
&=&
\prod_{s\in\eta}(a_\eta(s)+1+rl_\eta(s)),
\\
\label{h}
h_{\eta}(r)
&=&
\prod_{s\in\eta}(a_\eta(s)+r(l_\eta(s)+1)), \\
\left[ {0} \right]^r_\eta & = & 
\prod_{s\in\eta \atop{s \ne (1,1)}}(a'_\eta(s)-rl'_\eta(s)).
\end{eqnarray}

\subsection{Sutherland model and Jack polynomials}
Following Refs.\cite{arikawa00,arikawa99}, we formulate 
the dynamical spin structure factor $S^{zz}(Q,\omega)$ of the $1/r^2$ supersymmetric 
{\itshape t-J} model based on the freezing technique
\cite{polychronakos90,sutherland93}.

As an auxiliary, we introduce the Sutherland model \cite{sutherland1,sutherland2,sutherland3,sutherland4} with SU(2,1) supersymmetry \cite{katokuramotoSu}:
\begin{eqnarray}
{\mathcal H}_{CS} & = & -\frac{1}{2M} \sum_i \frac{\partial^2}{\partial x_i^2}
+ \frac{1}{M}\left( \frac{\pi}{L}\right)^2 \sum_{i<j}
\frac{\beta(\beta+\tilde{P}_{ij})}{\sin^2 \frac{\pi}{L}(x_i-x_j)}.
\end{eqnarray}
The system has $N_{\rm h}$ holes, $N_{\uparrow}$ up-spin electrons and $N_{\downarrow}$ down-spin ones, whose coordinates are represented by 
$x_i^{\rm h}$ for $i$-th hole, $x_i^\uparrow$ for $i$-th up-spin electron
and  $x_i^\downarrow$ for $i$-th down-spin electron. We arrange them as 
$x \equiv (x_1,x_2,\cdots,x_N)=(x_1^{\rm h},\cdots,x_{N_{\rm h}}^{\rm h},x_1^{\downarrow},\cdots,x_{N_\downarrow}^{\downarrow},
x_1^{\uparrow},\cdots,x_{N_\uparrow}^{\uparrow}) \equiv (x^{\rm h},x^{\downarrow},x^{\uparrow})$.
Here the graded exchange operator is defined as 
\begin{eqnarray}
\tilde{P}_{ij} = \sum_{\alpha,\beta} X_i^{\alpha\beta} X_j^{\beta \alpha} \theta_\beta,
\end{eqnarray}
where $X_j^{\beta\alpha}$ is the Hubbard operator which changes from state $\alpha$ to $\beta$ one at site $j$,
with $\alpha, \beta$ being either h (hole state), or one of $\sigma$ =$\uparrow, \downarrow$.
The sign factor $\theta_\beta$ is $-1$ if $\beta=$h and $1$ otherwise.
In order to reproduce the lattice model, we take the limit of large $\beta$ and $M$,
keeping the ratio $t=\beta/M$ fixed. 
Then the particles crystallize with equal distance from each other, and the resultant 
dynamics excluding phonons and uniform motion of the center of gravity is mapped to that of
the {\itshape t-J} model given by  Eq.(\ref{tj-hamiltonian}).
It can be shown that the intensity of the phonon correlation is smaller than the spin correlation
by a factor of ${\mathcal O}(\beta^{-1})$.
Here we take the lattice parameter $L/N$ as the unit of length.
Then we have the following relation
\begin{eqnarray}
{\mathcal H}_{CS} \rightarrow t \sum_{i<j} D_{ij}^{-2} \tilde{P}_{ij}.
\end{eqnarray}
For fixed numbers of $(N_{\rm h}, N_{\uparrow},N_{\downarrow})$, the right-hand side of the
above relation is the {\itshape t-J} model given by  Eq.(\ref{tj-hamiltonian})
with a trivial constant shift. 
Note that the symmetry of the wavefunction leads to the relation $s_{ij} \tilde{P}_{ij}=-1$, where $s_{ij}$ represents the exchange operator of the coordinates 
of particles $i$ and $j$. The interval $I=[1,N]$ denotes $\{i \in {\bf Z}| 1 \le i \le N \}$. We define $I_{\rm h}=[1,N_{\rm h}]$, $I_{\downarrow}=[N_{\rm h}+1,N_{\rm h}+N_{\downarrow}]$ and $I_{\uparrow}=[N_{\rm h}+N_{\downarrow}+1,N]$.  
The wave function of the ground state for a set of $(N_{\rm h},N_{\downarrow},N_{\uparrow})$ is given by
\begin{eqnarray}
\Psi_{\rm GS} & = & \prod_{i \ne j \in I}
\left( 1- \frac{z_j}{z_i} \right)^{\beta/2}
\prod_{\sigma=\uparrow,\downarrow} \prod_{i \ne j \in I_\sigma}
\left( 1- \frac{z_j}{z_i} \right)^{1/2},
\end{eqnarray}
where the complex coordinates $z=(z_1,\cdots,z_N)$ are related to 
the original ones $x=(x_1,\cdots,x_N)$ by $z_j = \exp (2\pi {\rm i} x_j/L)$.
The spectrum of the Sutherland model is conveniently analyzed with the use of a 
similarity transformation generated by 
\begin{equation}
{\mathcal O} =\prod_{i \ne j \in I}\left( 1- \frac{z_j}{z_i} \right)^{\beta/2} \prod_{i \in I} z_i^{(N_\uparrow-1)/2}.
\end{equation}
The transformed  Hamiltonian $\hat{\mathcal H} = {\mathcal O}^{-1} {\mathcal H}_{\rm CS} {\mathcal O}$ is 
\begin{eqnarray}
\hat{\mathcal H} & = & \frac{1}{2M} \left(\frac{2\pi}{L} \right)^2 \sum_{i=1}^N \left( \hat{d}_i + \beta \frac{N-1}{2} - \frac{N_\uparrow-1}{2} \right)^2. 
\end{eqnarray} 
Here $\hat{d}_i$ is called the Cherednik-Dunkl operator\cite{dunkl89,cherednik91} and is given by
\begin{eqnarray}
\hat{d}_i & = & z_i \frac{\partial}{\partial z_i} + \beta \sum_{k<i} 
\frac{z_i}{z_i-z_k}(1-s_{ik}) \nonumber \\
& & 
+ \beta \sum_{i<k} 
\frac{z_k}{z_i-z_k}(1-s_{ik}) + \beta(1-i).
\end{eqnarray}
It is known that $\hat{d}_i$ can be diagonalized simultaneously by homogeneous polynomials. 
In terms of the monomial $z^\nu = z_1^{\nu_1} \cdots z_N^{\nu_N}$,
the resultant eigenfunctions
$E_\eta(z;\beta)$ can be expressed as $E_\eta(z;\beta)=z^\eta+$ lower terms ({\itshape triangularity}),
and are called nonsymmetric Jack polynomials\cite{opdam95,macdonald97}. 
Here "lower terms" means a linear combination of the monomial $z^\nu$ such that $\nu \prec \eta$.
The eigenvalue $\bar{\eta}_i$ of $E_\eta(z;\beta)$  for $\hat{d}_i$ is given by $\bar{\eta}_i = \eta_i - \beta l'_\eta(i)$
for $i=1, \cdots ,N$.

\begin{figure}[ht]
%\begin{picture}(200,100)(-125,0)%
\begin{center}
\includegraphics[width=6cm]{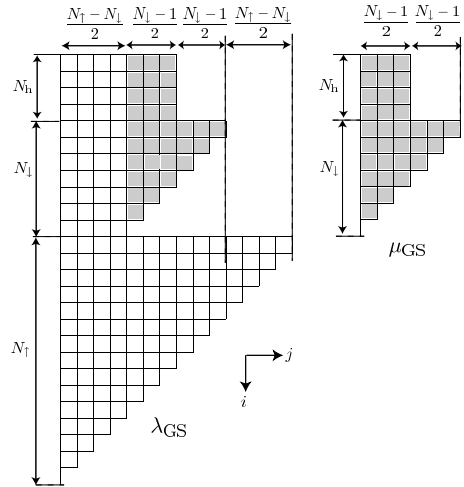}
\end{center}
%\end{picture}
\caption[]{
The diagrams of the ground state with $(N,N_{\rm h},N_{\downarrow},N_{\uparrow})
=(26,4,7,15)$.
$\lambda_{\rm GS}=(4^4,4^7,0^{15})+(3^4,\tilde{\delta}(7),\tilde{\delta}(15))$ (left)
and $\mu_{\rm GS}=(3^4,\tilde{\delta}(7))$ (right).
\label{gsconfig} 
}
\end{figure}

Since we are dealing with identical particles,
the eigenfunction should satisfy the following conditions of the SU(2,1) supersymmetry: \\
(i) symmetric with respect to the exchange between $z_i^{\rm h}$'s; \\
(ii)antisymmetric with respect to the exchange between $z_i^{\sigma}$'s with the same $\sigma$. \\
By taking a linear combination of $E_\lambda(z;\beta)$, we can construct a polynomial 
$K_\lambda(z;\beta)$ with SU(2,1) supersymmetry\cite{kato98,baker97,dunkl98,baker00}. 
The above {\itshape triangular} structure of $E_\lambda(z;\beta)$ is inherited to $K_\lambda(z;\beta)$.
We specify the set of momenta as $\lambda=(\lambda^{\rm h},\lambda^{\downarrow},\lambda^{\uparrow}) \in \Lambda_N$, where $\lambda^{\rm h} \in \Lambda^+_{N_{\rm h}}$
and $\lambda^{\sigma} \in \Lambda^{+,>}_{N_{\sigma}}$ ($\sigma = \uparrow,\downarrow$).
For the ground state, we have
$\lambda=\lambda_{\rm GS}=(\lambda_{\rm GS}^{\rm h},\lambda_{\rm GS}^{\downarrow},\lambda_{\rm GS}^{\uparrow})$ with $\lambda_{\rm GS}^{\rm h}=(\frac{N_\uparrow-1}{2})^{N_{\rm h}}$,
$\lambda_{\rm GS}^\downarrow =(\tilde{\delta}(N_\downarrow) +
(\frac{N_{\uparrow}-N_\downarrow}{2})^{N_\downarrow})$ and
$\lambda_{\rm GS}^{\uparrow}=\tilde{\delta}(N_\uparrow)$ (see Figure \ref{gsconfig}).
$K_\lambda(z;\beta)$ is normalized so that the coefficient of the monomial $z^\lambda$ is 
unity.

We define the inner product of functions $f(z)$ and $g(z)$ in 
$n$ complex variables,
$z=(z_1,z_2,\cdots,z_n)$ as follows:
\begin{eqnarray}
\langle f,g \rangle_n^\beta & = & \prod_{j=1}^n \oint_{|z_j|=1} \frac{d z_j}{2\pi {\rm i}z_j}
\overline{f(z)}g(z) \prod_{1\le k<l \le n}|z_k -z_l|^{2\beta}
\end{eqnarray}
where $\overline{f(z)}$ denotes the complex conjugation of $f(z)$.
We give some examples of the SU(2,1) Jack polynomials. For the case of $N_{\rm h}=N$,
the polynomial $K_\lambda(z;\beta)$ ($\lambda \in \Lambda_N^+$) reduces to the 
(monic) Jack polynomial $P_\lambda(z,\beta)$. The antisymmetric Jack polynomial
$P_\lambda(z,\beta+1) \prod_{i<j \in I} (z_i-z_j)$ is given by $K_{\lambda+\tilde{\delta}(N)}(z,\beta)$ ($\lambda \in \Lambda_N^+$) with $(N_{\rm h},N_\downarrow)=(0,0)$. 
The SU(2,1) Jack polynomials are orthogonal with respect to the above inner product\cite{baker00,takemura97}:
\begin{eqnarray}
\hspace{-1cm}
\langle K_\lambda,K_\mu \rangle^\beta_N
& = & 
\delta_{\lambda \mu}
\frac{\Gamma[N \beta +1]}{\Gamma[\beta+1]^N}
\frac{N_{\rm h}! N_{\uparrow}! N_{\downarrow}!}{\rho_\lambda(\beta)}
\frac{d_\lambda'(\beta)}{d_\lambda(\beta)}
\frac{f_\lambda(\beta;1+\beta N)}{f_\lambda(\beta;1+\beta (N-1))},
\end{eqnarray}
for compositions 
$\lambda,\mu \in \Lambda_{N_{\rm h}}^+ \times \Lambda_{N_{\downarrow}}^{+,>}
\times \Lambda_{N_{\uparrow}}^{+,>} \subset \Lambda_N$. 
Here $\rho_\lambda(\beta)$ is given by the product $\rho_\lambda(\beta) = \rho^{\rm h}_\lambda(\beta)
\rho^{\uparrow}_\lambda(\beta) \rho^{\downarrow}_\lambda(\beta)$ with
\begin{eqnarray}
\rho_\lambda^{\rm h}(\beta) & = & \prod_{i<j \in I_{\rm h}} 
\frac{\bar{\lambda}_i-\bar{\lambda}_j+\beta}{\bar{\lambda}_i-\bar{\lambda}_j}, \\
\rho_\lambda^{\sigma}(\beta) & = & \prod_{i<j \in I_{\sigma}} 
\frac{\bar{\lambda}_i-\bar{\lambda}_j-\beta}{\bar{\lambda}_i-\bar{\lambda}_j} \
\ (\sigma=\uparrow,\downarrow).
\end{eqnarray}

\begin{figure}[ht]
%\begin{picture}(200,100)(-125,0)%
\begin{center}
\includegraphics[width=6cm]{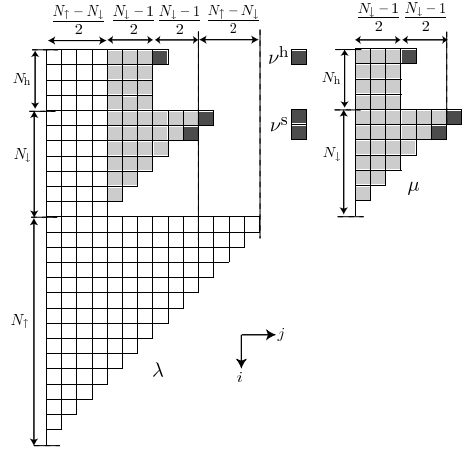}
\end{center}
%\end{picture}
\caption[]{
The diagrams that appear in the expansions (\ref{f:expansioncoef1}) and  (\ref{f:expansioncoef2}) 
under the conditions (\ref{f:small1}) and (\ref{f:small2}) (the small momentum region).
These diagrams correspond to the case with 
$(N,N_{\rm h},N_{\downarrow},N_{\uparrow})
=(26,4,7,15)$. 
$\lambda=\lambda_{\rm GS}+(\nu^{\rm h},\nu^{\rm s},0^{N_\uparrow})$ (left) and
$\mu=\mu_{\rm GS}+(\nu^{\rm h},\nu^{\rm s})$ (right)
with $\nu^{\rm h}=(1,0^3)$ and  
$\nu^{\rm s}=(1^2,0^5)$.
The diagrams $\lambda_{\rm GS}$ and $\mu_{\rm GS}$ are shown in Figure \ref{gsconfig}.  
The diagrams shown in this figure do not contribute to $S^{zz}(Q,\omega)$,
owing to $\nu^{\rm h} = (1,0^{N_{\rm h}-1})$ (see Eq.(\ref{expa2})).
\label{exconfig} 
}
\end{figure}
The operators $n^{\rm h}_Q
=\sum_j X_j^{00} e^{{\rm i} Q j}/\sqrt{N}$ and
$n^{\sigma}_Q =\sum_j X_j^{\sigma\sigma} e^{{\rm i} Q j}/\sqrt{N}$ 
can be expressed for $Q=2\pi m/N$ as
$n^{\rm h}_Q = p^{\rm h}_{m}/\sqrt{N}$ and 
$n^{\sigma}_Q = p^{\sigma}_{m}/\sqrt{N}$, respectively.    
Here we have introduced power sums $p_m^{\alpha}= \sum_{i \in I_{\alpha}} z_i^m$  ($\alpha={\rm h},\uparrow,\downarrow$). 
In the lattice model we have the completeness relation
$\sum_{\sigma=\uparrow,\downarrow} X_i^{\sigma\sigma} + X_i^{00}=1$. Therefore, 
in order to calculate $S^{zz}(Q,\omega)$, we need to know two types of the
 expansion coefficients, $c_{\lambda}^{\rm h}$ and  $c_{\lambda}^{\downarrow}$, which appear in
\begin{eqnarray}
p_m^{\rm h} K_{\lambda_{\rm GS}} & = & \sum_{\lambda} c_\lambda^{\rm h} K_\lambda(z;\beta), 
\label{f:expansioncoef1} \\ 
p_m^{\downarrow} K_{\lambda_{\rm GS}} & = & \sum_{\lambda} c_\lambda^{\downarrow} K_\lambda(z;\beta).\label{f:expansioncoef2}
\end{eqnarray} 
Using these coefficients, in the lattice limit ($\beta \rightarrow \infty$), the spin operator $S_j^z$ 
can be expressed by
$S_j^z = (X_j^{\uparrow\uparrow} -X_j^{\downarrow\downarrow})/2 = 
1/2 - X_j^{00}/2- X_j^{\downarrow\downarrow}$. Therefore for $Q > 0$, we have
the following relation
\begin{eqnarray}
S^{zz}(Q,\omega) & = & 
\frac{1}{N}
\sum_\lambda \left(c_\lambda^{\downarrow}+\frac{c_\lambda^{\rm h}}{2} \right)^2
\frac{\langle K_\lambda, K_\lambda \rangle_N}{\langle K_{\lambda_{\rm GS}}, K_{\lambda_{\rm GS}} \rangle_N} \delta(\omega -\Delta E_\lambda).
\label{f:starting}
\end{eqnarray}
For $Q=0$, there is a finite intensity only at $\omega=0$,
which is given by $N \bar{m}^2/4$.
It is difficult to derive $c_{\lambda}^{\rm h}$ and  $c_{\lambda}^{\downarrow}$ for general values of $Q=2\pi m/N$.
In the next section, we show that there occurs a drastic simplification in the small momentum region.  

\section{Matrix element}
\label{sec.matrixelement}
In the small momentum region, we have some special properties 
that the calculation for the expansion coefficient and norm can be essentially reduced to those 
of symmetric Jack polynomials~\cite{arikawa04b}.
First we summarize the necessary formula to evaluate the expansion coefficients 
$c_\lambda^{\downarrow}$ and $c_\lambda^{\rm h}$.
Next we obtain the expansion coefficients  
by use of the replica type technique \cite{gangardt01}. 

\subsection{Small momentum region}
We consider the case where the following two
conditions are satisfied: 
\begin{eqnarray}
m  &\le & (N_\uparrow-N_\downarrow)/2, \label{f:small1} \\
m  &\le & (N_\downarrow-1)/2. \label{f:small2}
\end{eqnarray}
(Note that these conditions constitute the small momentum region.)
In this region,  
owing to the {\itshape triangular} structure of the polynomial $K_\lambda(z,\beta)$, the composition $\lambda$ contributing to the summation (\ref{f:starting})
is restricted to the form $\lambda=\lambda_{\rm GS}+(\nu^{\rm h},\nu^{\rm s},0^{N_\uparrow})$ (see Figures \ref{gsconfig} and \ref{exconfig}).
We define $\mu_{\rm GS}=(\mu_{\rm GS}^{\rm h},\mu_{\rm GS}^{\downarrow})$ with 
$\mu_{\rm GS}^{\rm h}=((\frac{N_\downarrow-1}{2})^{N_{\rm h}})$ and $\mu_{\rm GS}^\downarrow=\tilde{\delta}(N_\downarrow)$.
Under the conditions (\ref{f:small1}) and (\ref{f:small2}), by extending a calculation by Baker and Forrester \cite{baker97}, we obtain the following relation:
\begin{eqnarray}
K_\lambda(z;\beta) & = & \tilde{K}_\mu(\tilde{z};\beta')  \prod_{j \in I_{\rm h} \cup I_\downarrow} z_j^{\frac{N_\uparrow-N_\downarrow}{2}} 
\prod_{i<j \in I_\uparrow}(z_i-z_j).
\label{f:baker}
\end{eqnarray} 
where $\beta'=\beta/(\beta+1)$, $\mu=(\mu^{\rm h}_{\rm GS},
\mu^{\downarrow}_{\rm GS})+(\nu^{\rm h},\nu^{\rm s}) \in \Lambda_{N_{\rm h}}^+
\times \Lambda_{N_{\downarrow}}^{+,>}$ and $\tilde{z}=(z^{\rm h},z^{\downarrow})$. 
Here 
$\tilde{K}_\mu(\tilde{z};\beta')$
is a Jack polynomial with SU(1,1) supersymmetry, which is a 
linear combination of $N_{\rm h}+N_{\downarrow}$ 
variables non-symmetric Jack polynomials
$E_\eta(\tilde{z};\beta')$. 
The polynomial $E_\eta(\tilde{z};\beta')$ can be obtained  by substitution 
$z_j=0$ for $j \in [N_{\rm h}+N_\downarrow+1,N]$ in 
$N$ variables non-symmetric Jack polynomial $E_\eta(z;\beta')$. 
The SU(1,1) Jack polynomial $\tilde{K}_\mu(\tilde{z};\beta')$ is symmetric with respect to
the
exchange between $z_i^{\rm h}$'s and  antisymmetric with respect to 
the
exchange between $z_i^{\downarrow}$'s. 
For the composition $\mu_{\rm GS}$, the SU(1,1) Jack polynomial
$\tilde{K}_{\mu_{\rm GS}}(\tilde{z},\beta')$ is independent of the parameter $\beta'$, and 
is explicitly given by
\begin{eqnarray}
\tilde{K}_{\mu_{\rm GS}}(\tilde{z},\beta') & = & \prod_{j \in I_{\rm h}} z_j^{(N_\downarrow-1)/2}
\prod_{i<j \in I_\downarrow} (z_i-z_j).
\end{eqnarray}
Under the conditions (\ref{f:small1}) and (\ref{f:small2}), 
the norm of the above states can be reduced to the following form\cite{arikawa04b}:
\begin{eqnarray}
\frac{\langle K_\lambda, K_\lambda \rangle^\beta_N}
{\langle K_{\lambda_{\rm GS}}, K_{\lambda_{\rm GS}} \rangle^\beta_N}
& = & 
\frac{\langle \tilde{K}_\mu, \tilde{K}_\mu \rangle^{\beta'}_{N_{\rm h}+N_\downarrow}}
{\langle \tilde{K}_{\mu_{\rm GS}}, \tilde{K}_{\mu_{\rm GS}} \rangle^{\beta'}_{N_{\rm h}+N_\downarrow}} \label{f:normsimple1} \\
& = &
\frac{d'_{\nu^{\rm h}}(\beta'')}{h_{\nu^{\rm h}}(\beta'')}
\frac{f_{\nu^{\rm h}}(\beta''; \beta'' N_{\rm h})}{f_{\nu^{\rm h}}
(\beta'';1+\beta''(N_{\rm h}-1))} \nonumber \\
& & 
\times 
\frac{d'_{\nu^{\rm s}}(\bar{\beta}')}{h_{\nu^{\rm s}}(\bar{\beta}')}
\frac{f_{\nu^{\rm s}}(\bar{\beta}'; \bar{\beta}' (N_{\downarrow}+\beta'' N_{\rm h}))}{f_{\nu^{\rm s}}
(\bar{\beta}';1+\bar{\beta}'(N_{\downarrow}+\beta'' N_{\rm h}-1))} \nonumber,
\end{eqnarray}
where $\beta''=\beta'/(\beta'+1) = \beta/(2\beta+1)$ and $\bar{\beta'} = \beta'+1 = (2\beta+1)/(\beta+1)$.
We remark that for monic symmetric Jack polynomials $P_\lambda(z,\beta)$, the following relation is obtained\cite{macdonald95}:
\begin{eqnarray}
\frac{\langle P_\lambda, P_\lambda \rangle^\beta_N}
{\langle 1 , 1 \rangle^\beta_N}
& = & 
\frac{d'_{\lambda}(\beta)}{h_{\lambda}(\beta)}
\frac{f_{\lambda}(\beta; \beta N)}{f_{\lambda}
(\beta;1+\beta(N-1))}.
\end{eqnarray}
The relations (\ref{f:baker}) 
and (\ref{f:normsimple1})
hold if both the conditions $0 \le \lambda_j \le N_\uparrow-1$ for $j \in I_{\rm h} \cup I_\downarrow$ and 
$\lambda^\uparrow=\lambda^{\uparrow}_{\rm GS}$ 
are satisfied. However if Eqs.(\ref{f:small1}) and (\ref{f:small2}) are
not satisfied, the $K_\lambda$ without satisfying Eq.(\ref{f:baker}) may involve 
in the summations (\ref{f:expansioncoef1}) and  (\ref{f:expansioncoef2}).
In the strong coupling limit ($\beta \rightarrow \infty$), the parameter $\beta'=\beta/(\beta+1)$ approaches 
unity.
In this limit, the eigenstates satisfying the above conditions "$0 \le \lambda_j \le N_\uparrow-1$ for $j \in I_{\rm h} \cup I_\downarrow$ and $\lambda^\uparrow=\lambda^{\uparrow}_{\rm GS}$" 
are SU(2,1) Yangian highest weight states (YHWS) \cite{ha94} in the $1/r^2$ supersymmetric {\itshape t-J} model, which 
can be mapped to that of the SU(1,1) Sutherland model 
with coupling parameter $\beta=1$
\cite{saiga96}.
As employed in Refs.~\cite{arikawa04b,arikawa01,kato98a},
if the excited states are restricted 
within the YHWS, the derivation of the correlation functions
can be reduced to that for the Sutherland model with  SU(1,1) supersymmetry.   
We would like to stress that if the small momentum conditions (\ref{f:small1}) and 
(\ref{f:small2}) are not satisfied, non-YHWS may contribute in the 
excited states of 
$S^{zz}(Q,\omega)$.

\subsection{Replica type technique}
Using the replica type technique \cite{gangardt01}, 
we derive the analytic formula of the coefficients 
$c_\lambda^{\rm h}$ and $c_\lambda^{\downarrow}$.  
For indices $\alpha = {\rm h}$ and $\downarrow$, we define the quantity ${\mathcal Z}^{\alpha}(\theta)$  as follows:
\begin{eqnarray}
{\mathcal Z}^{\alpha}(\theta) & = & \prod_{j \in I_{\alpha}} (1-e^{-{\rm i}\theta} z_j).
\end{eqnarray}
For any 
real
parameter $u$, we have the following relation: 
\begin{eqnarray*}
\hspace{-1cm}
\left({\mathcal Z}^{\alpha}(\theta) \right)^u
& = & \exp\left(u \sum_{j \in I_\alpha}\ln(1-e^{-{\rm i} \theta} z_j) \right)  
 =  \exp\left( -u  \sum_{m=1}^\infty e^{-{\rm i} m \theta} \frac{p_m^{\alpha}}{m} \right).
\end{eqnarray*}
By use of the above relation,
the expansion coefficients $c_{\lambda}^{\alpha}$ ($\alpha ={\rm h}, \downarrow$) in Eqs.(\ref{f:expansioncoef1}) and (\ref{f:expansioncoef2}) are given by
\begin{eqnarray}
\hspace{-1cm}
c_{\lambda}^{\alpha} & = & \frac{\langle {K}_\lambda, (p_m^{\rm \alpha} \times 
{K}_{\lambda_{\rm GS}})\rangle^{\beta}_N}
{ \langle {K}_\lambda, {K}_\lambda \rangle^{\beta}_N } 
= \left. \lim_{u \rightarrow 0} \frac{1}{\rm i} \frac{1}{u} \frac{\partial}{\partial \theta}
\frac{\langle {K}_\lambda, 
\left({\mathcal Z}^{\alpha}(\theta)\right)^u {K}_{\lambda_{\rm GS}}   
\rangle^{\beta}_N}
{\langle {K}_\lambda, {K}_\lambda \rangle^{\beta}_{N}}\right|_{\theta=0} ,
\label{eq:replicares21}
\end{eqnarray}
where momentum conservation $|| \lambda || - || \lambda_{\rm  GS}|| =m$ is satisfied.
We introduce the similar type of expansions for the SU(1,1) Jack polynomials:
\begin{eqnarray}
p_m^{\rm h} \tilde{K}_{\mu_{\rm GS}}(\tilde{z},\beta') & = & \sum_{\mu} \tilde{c}_\mu^{\rm h} \tilde{K}_\mu(\tilde{z},\beta'), 
\label{f:expansioncoef3} \\ 
p_m^{\downarrow} \tilde{K}_{\mu_{\rm GS}}(\tilde{z},\beta') & = & \sum_{\mu} \tilde{c}_\mu^{\downarrow} \tilde{K}_\mu(\tilde{z},\beta').\label{f:expansioncoef4}
\end{eqnarray}
In a similar manner, the expansion coefficients of $\tilde{c}_{\mu}^{\alpha}$ ($\alpha={\rm h},\downarrow$) in Eqs.(\ref{f:expansioncoef3}) and  (\ref{f:expansioncoef4}) are given by 
\begin{eqnarray}
\tilde{c}_{\mu}^{\alpha} & = & 
\left. \lim_{u \rightarrow 0} \frac{1}{\rm i} \frac{1}{u} \frac{\partial}{\partial \theta}
\frac{\langle \tilde{K}_\mu, 
\left({\mathcal Z}^{\alpha}(\theta)\right)^u \tilde{K}_{\mu_{\rm GS}}   
\rangle^{\beta'}_{N_{\rm h}+N_\downarrow}}
{\langle \tilde{K}_\mu, \tilde{K}_\mu \rangle^{\beta'}_{N_{\rm h}+N_\downarrow}} \right|_{\theta=0}, 
\label{eq:replicares11}
\end{eqnarray}
where $|| \mu || - || \mu_{\rm  GS}|| =m$.

Under the conditions (\ref{f:small1}) and (\ref{f:small2}), by use of Eq.(\ref{f:baker}), for given compositions 
$\lambda=\lambda_{\rm GS}+(\nu^{\rm h},\nu^{\rm s},0^{N_\uparrow})$ and  $\mu=\mu_{\rm GS}+(\nu^{\rm h},\nu^{\rm s})$
\footnote{The $\nu^{\rm h}$ and $\nu^{\rm s}$ are the partitions with $\nu^{\rm h} \in \Lambda^+_{N_{\rm h}}$
and $\nu^{\rm s} \in \Lambda^+_{N_{\downarrow}}$, respectively.
}, 
we can show the relation
\footnote{In the strong coupling limit $\beta \rightarrow \infty$, we have 
$\beta' \rightarrow 1$. For YHWS, Eqs.~(\ref{f:normsimple1}) and (\ref{eq:guarantee})  
reflect the equivalence between the freezing approach and mapping of the eigenstates of 
the $1/r^2$ supersymmetric {\itshape t-J} model into those of the SU(1,1) Sutherland model.}
\begin{equation}
c_\lambda^\alpha = \tilde{c}_\mu^\alpha. \qquad (\alpha = {\rm h}, \downarrow)
\label{eq:guarantee}
\end{equation}
This relation means that
the expansion coefficients for the SU(2,1) Jack polynomials
can be expressed by those for the SU(1,1) Jack polynomials,
provided the conditions (\ref{f:small1}) and (\ref{f:small2}) are satisfied.
%%%%%%%%%%%%%

%%%
Next, for given parameters $(p,q)$, we consider the following expansion: 
\begin{eqnarray}
\left({\mathcal Z}^{\rm h}(0)\right)^p
\left({\mathcal Z}^{\downarrow}(0)\right)^q \tilde{K}_{\mu_{\rm GS}}(\tilde{z},\beta') 
& = & \sum_\mu  
\chi_\mu(\beta') \tilde{K}_\mu(\tilde{z};\beta').
\end{eqnarray}
The formula of the expansion coefficient $\chi_\mu(\beta')$ for arbitrary $\mu$ has not been proved yet.
However, in the small momentum region, we have a formula for $\chi_\mu(\beta')$. For $\mu=\mu_{\rm GS}+
(\nu^{\rm h},\nu^{\rm s})$, we have the following relation \cite{arikawa04b}:
\begin{eqnarray}
\chi_\mu(\beta') & = & 
\frac{f_{\nu^{\rm h}}(\beta''; -p+\beta' q)}{d'_{\nu^{\rm h}}(\beta'')}
\times 
\frac{f_{\nu^{\rm s}}(\bar{\beta}'; -q)}{d'_{\nu^{\rm s}}(\bar{\beta}')}.
\label{separatedexpand}
\end{eqnarray}    
By use of Eqs.(\ref{eq:replicares11}) and (\ref{separatedexpand}) with $(p,q)=(0,u)$,
we can obtain the coefficient $c_\lambda^{\downarrow}$.
For $\lambda=\lambda_{\rm GS}+(\nu^{\rm h},\nu^{\rm s},0^{N_\uparrow})$,
it is given by
\begin{eqnarray}
c_\lambda^{\downarrow} 
& = & -\beta'' m \frac{\left[ 0 \right]^{\beta''}_{\nu^{\rm h}} \times f_{\nu^{\rm s}}(\bar{\beta}';0)}
{d'_{\nu^{\rm h}}(\beta'') d'_{\nu^{\rm s}}(\bar{\beta}') } 
+ m
\frac{ f_{\nu^{\rm h}}(\beta'';0) \times \left[ 0 \right]^{\bar{\beta}'}_{\nu^{\rm s}} }
{d'_{\nu^{\rm h}}(\beta'') d'_{\nu^{\rm s}}(\bar{\beta}') }, 
\label{expa}
\end{eqnarray}
where $m=||\nu^{\rm h}||+||\nu^{\rm s}||$.
In a similar manner, we can derive the coefficient $c_{\lambda}^{\rm h}$ 
(see Eq.(\ref{eq:expandhole})). 
The first term in Eq.(\ref{expa}) can then be represented by $-\beta'' c_\lambda^{\rm h}$, 
which vanishes unless $\nu^{\rm s} = (0^{N_\downarrow})$.
We consider the quantity $c_\lambda= c_\lambda^{\downarrow} + \beta'' c_\lambda^{\rm h}$,
which is nothing but the second term of  
the right-hand side in Eq.(\ref{expa}).
Owing to the factor $f_{\nu^{\rm h}}(\beta'';0)$, the $c_\lambda$ vanishes unless $\nu^{\rm h} = (0^{N_{\rm h}})$.
Therefore, the $c_\lambda$ can be simplified as 
\begin{eqnarray}
c_\lambda & = & ||\nu^{\rm s}|| 
\frac{\left[ 0 \right]^{\bar{\beta}'}_{\nu^{\rm s}} }
{d'_{\nu^{\rm s}}(\bar{\beta}') }
\times \delta_{(\nu^{\rm h}),(0^{N_{\rm h}})}. 
\label{expa2}
\end{eqnarray}
We notice that while $c_\lambda^{\downarrow}$ is related to the expansion coefficients for the SU(1,1) Jack polynomials,
the quantity $c_\lambda$ given by Eq.(\ref{expa2}) is related to those
 for the monic symmetric Jack polynomials.
In fact, for the monic symmetric Jack polynomials,
one has the following formula \cite{hanlon92}:
\begin{eqnarray}
\sum_{j=1}^n z_j^m & = & m \times \sum_{\lambda \in \Lambda_n^+ \atop{||\lambda||=m}}
\frac{\left[ 0 \right]^{{\beta}}_{\lambda} }
{d'_{\lambda}(\beta) } \ P_\lambda(z;\beta).
\label{hanlonformula}
\end{eqnarray}
In the strong coupling limit ($\beta \rightarrow \infty$), we have ${\beta}'' \rightarrow 1/2$,
and therefore the $c_\lambda$ 
approaches $c_\lambda^{\downarrow} + c_\lambda^{\rm h}/2$.
This is precisely the quantity which appears in $S^{zz}(Q,\omega)$
(see Eq.(\ref{f:starting})).
Thus, no charge excitation contributes to $S^{zz}(Q,\omega)$ in the small momentum region.
Since we have $\bar{\beta'} \rightarrow 2$ in the limit $\beta \rightarrow \infty$,
the coefficient $c_\lambda$ vanishes 
owing to the factor $\left[ 0 \right]^{\bar{\beta}'}_{\nu^{\rm s}}$, 
if partition $\nu^{\rm s}$ 
contains $s=(i,j)=(2,3)$. Then, $S^{zz}(Q,\omega)$ is determined by three parameters 
$(\lambda_{\rm a},\lambda_{\rm s1},\lambda_{\rm s2})$. 
Partition $\nu^{\rm s}$ is restricted to 
$\nu^{\rm s}=(\lambda_{\rm a}, 2^{\lambda_{\rm s2}-1},1^{\lambda_{\rm s1}-\lambda_{s2}},
0^{N_\downarrow-\lambda_{\rm s1}})$.  
These three parameters are related directly with 
momenta of the 
elementary excitations: one antispinon and two spinons.

\begin{figure}[ht]
%\begin{picture}(200,100)(-125,0)%
\begin{center}
\includegraphics[width=6cm]{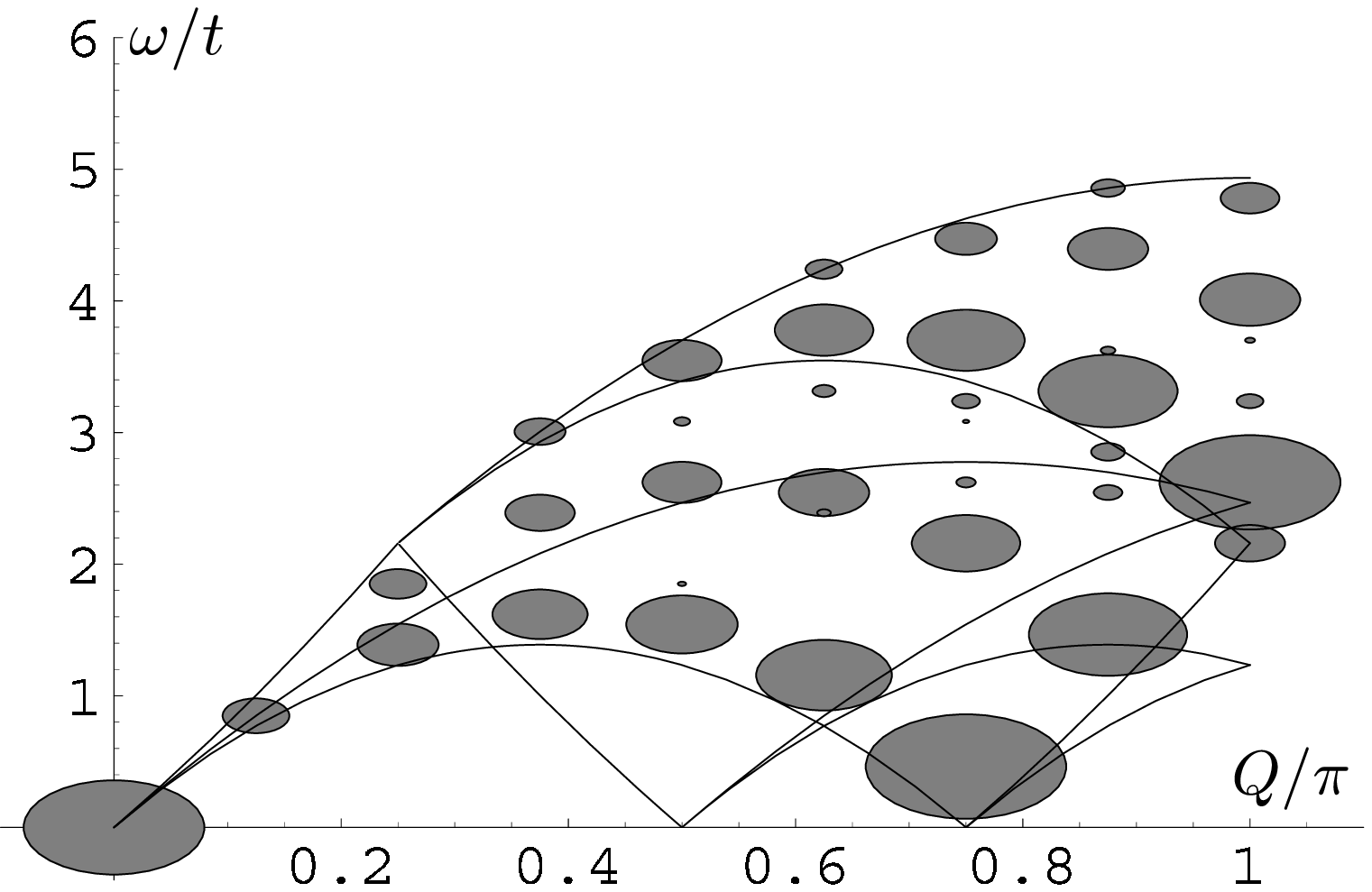}
\includegraphics[width=6cm]{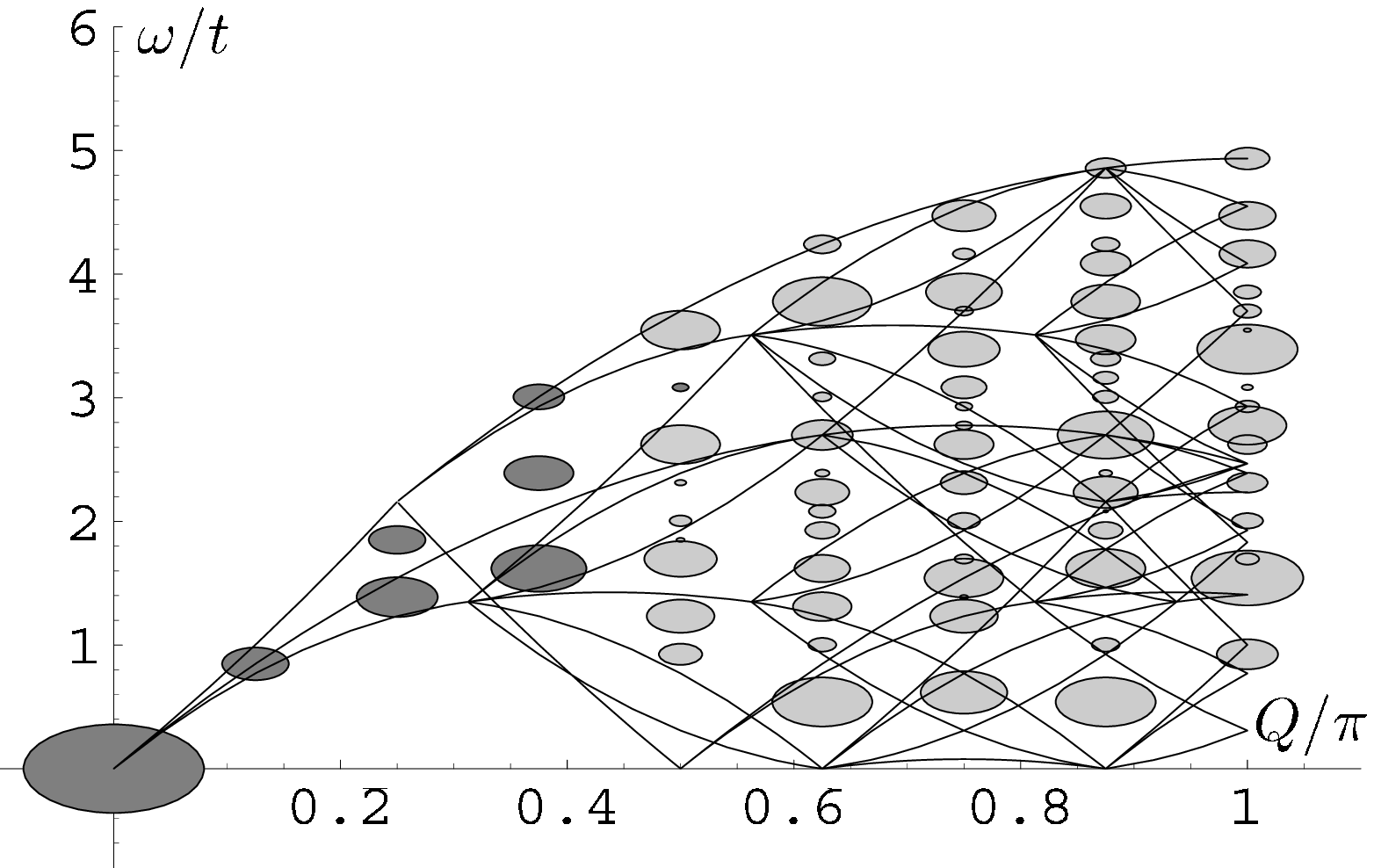}
\end{center}
%\end{picture}
\caption[]{
$S^{zz}(Q,\omega)$ for $(N,N_{\rm h},N_\uparrow,N_\downarrow)=(16,0,10,6)$
(left) and
$(16,2,9,5)$ (right). Each spectral weight is proportional to the area of the oval. 
For dark shaded ovals in the right figure,
excitation energies and spectral weights agree with
those in the Haldane-Shastry model (the left figure), which can be expressed by Eq.(\ref{eq:finiteszzexp1}) and (\ref{dsip}). 
\label{comp.fig1}
}
\end{figure}

\begin{figure}[ht]
%\begin{picture}(200,100)(-125,0)%
\begin{center}
\includegraphics[width=6cm]{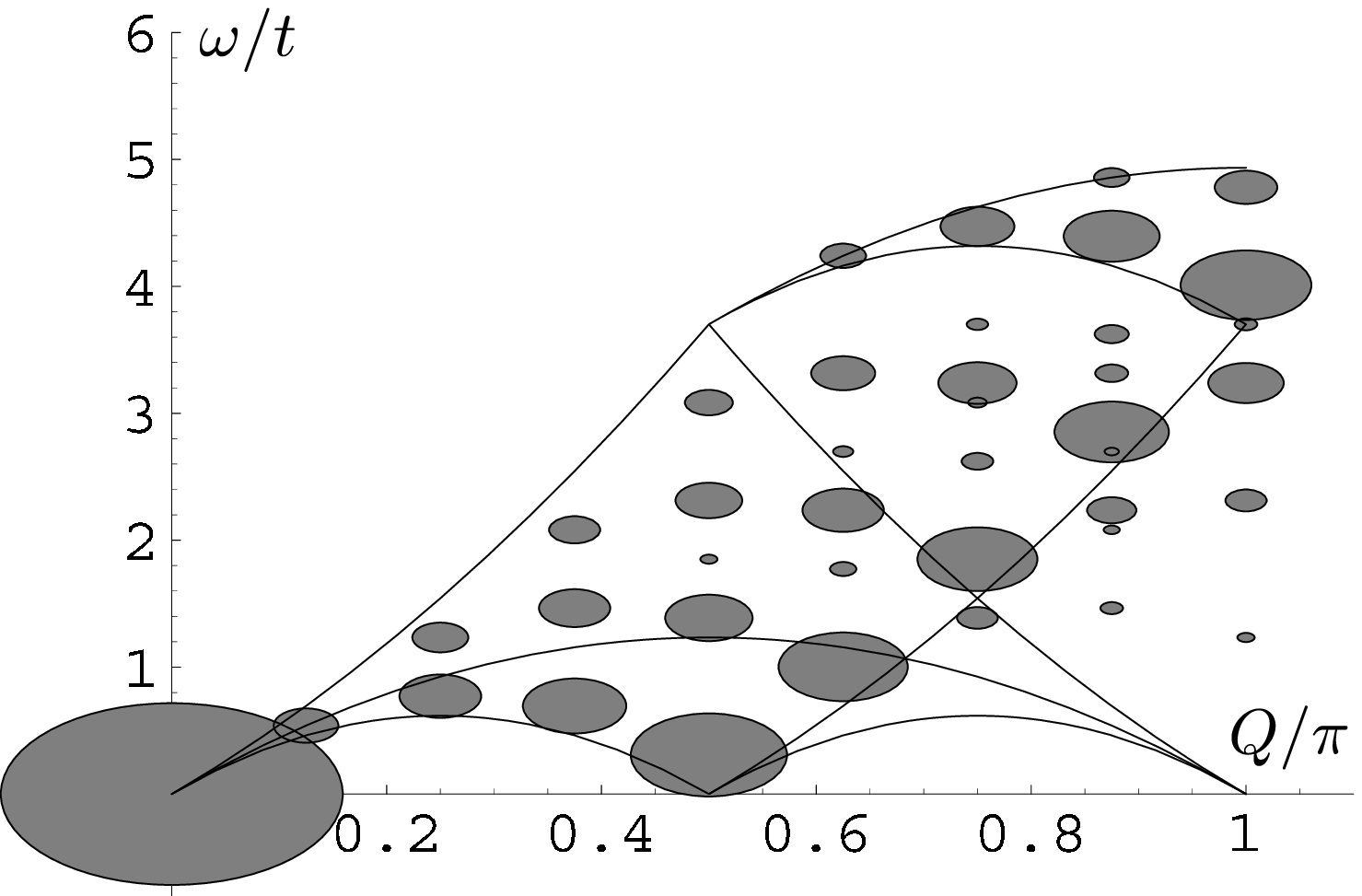}
\includegraphics[width=6cm]{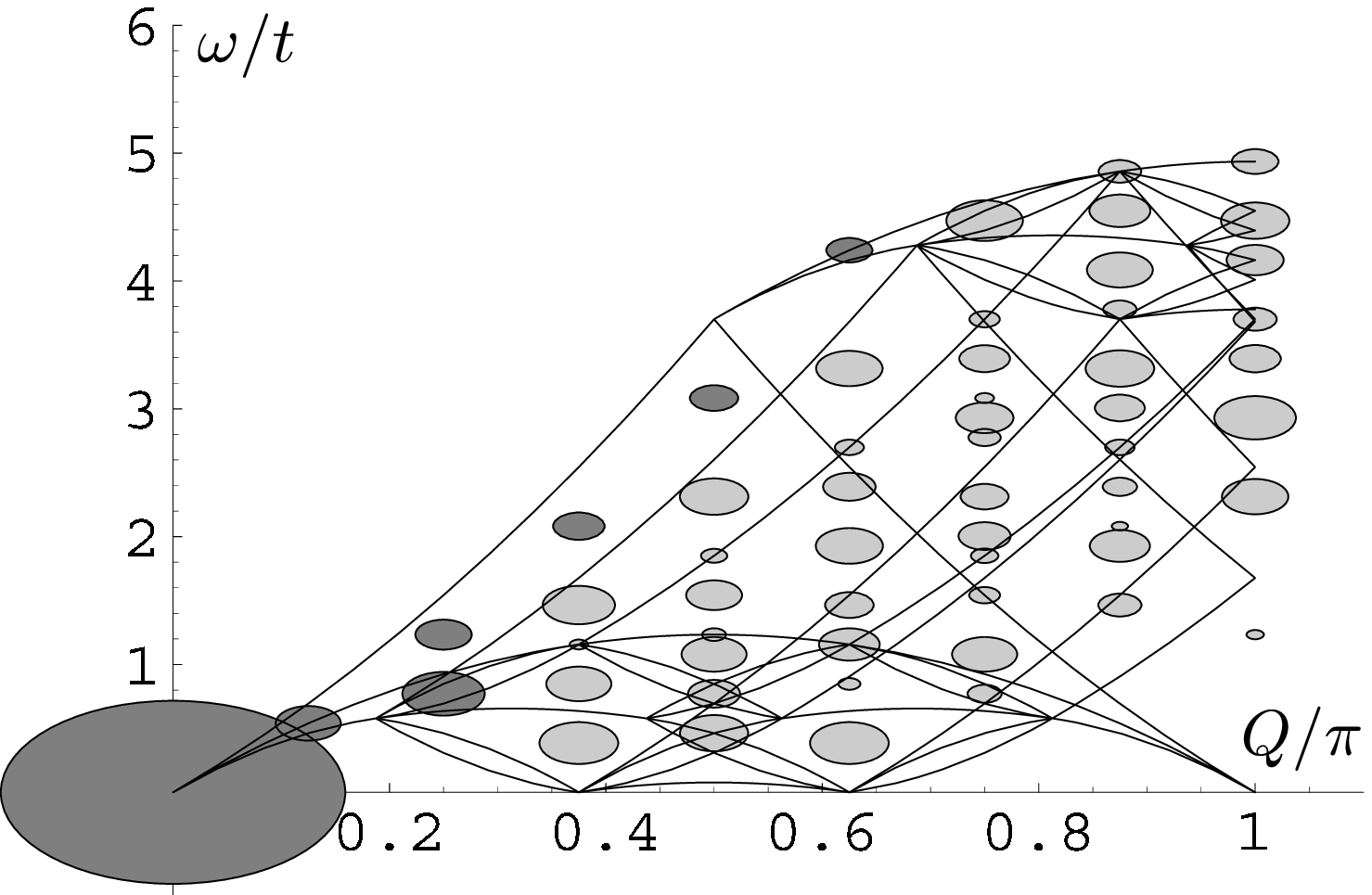}
\end{center}
%\end{picture}
\caption[]{
$S^{zz}(Q,\omega)$ for $(N,N_{\rm h},N_\uparrow, N_\downarrow)=(16,0,12,4)$ (left) and
$(16,2,11,3)$ (right). For dark shaded ovals in the right figure,
excitation energies and spectral weights agree with
those in the Haldane-Shastry model (the left figure), which can be expressed by Eq.(\ref{eq:finiteszzexp1}) and (\ref{dsip}). 
\label{comp.fig2}
}
\end{figure}

\section{Results}
\label{sec.result}
Using the relations (\ref{f:normsimple1}) and (\ref{expa2}), we can express the $S^{zz}(Q,\omega)$ in terms of three parameters 
$(\lambda_{\rm a},\lambda_{\rm s1},\lambda_{\rm s2})$.
In the small momentum region (see Eqs.(\ref{f:small1}) and (\ref{f:small2})), we obtain the following expression:
\begin{eqnarray}
S^{zz}(Q,\omega)
&=&
\frac{m^2}{2N}
\sum_{\lambda_{{\rm s1}} \ge \lambda_{{\rm s2}},\lambda_{{\rm a}}}
\delta_{m, || \nu^{\rm s} ||}
\delta(\omega-\Delta E_{\lambda}) \nonumber \\
& & \times
(\lambda_{\rm a}-1) (\lambda_{\rm a}+N-(N_\uparrow-N_\downarrow)-1) 
(\lambda_{{\rm s}1}-\lambda_{{\rm s}2}+\frac{1}{2})
\nonumber \\
& & \times
\prod_{j=1}^2 
\frac{1}{(\lambda_{\rm a}+2\lambda_{{\rm s}j}-j-1) (\lambda_{\rm a}+2\lambda_{{\rm s}j}-j) }
\nonumber \\
& & \times
\prod_{j=1}^2
\frac{\Gamma[\lambda_{{\rm s}j}-\frac{j-1}{2}] 
\Gamma[\frac{N}{2}-\frac{N_\uparrow-N_\downarrow}{2}-\lambda_{{\rm s}j}+\frac{j}{2}]}
{\Gamma[\lambda_{{\rm s}j}-\frac{j-2}{2}] 
\Gamma[\frac{N}{2}-\frac{N_\uparrow-N_\downarrow}{2}
-\lambda_{{\rm s}j}+\frac{j+1}{2}]},
\label{eq:finiteszzexp1}
\end{eqnarray}
where $Q=2\pi m/N$ and $|| \nu^{\rm s}  || = 
\lambda_{\rm a}+\sum_{j=1}^2(\lambda_{{\rm s}j}-1)$. The excitation 
energy $\Delta E_\lambda$ is given by
\begin{eqnarray}
\Delta E_{\lambda}/t & = &
\left( \frac{2 \pi}{N} \right)^2
\left[
\left(\lambda_{\rm a}-2 \right)
\left(\frac{N}{2}-\frac{N_\uparrow-N_\downarrow}{2}+\frac{\lambda_{\rm a}}{2} \right) 
\right. \nonumber \\
& & \left.
+ \sum_{j=1}^2 \lambda_{{\rm s}j} \left( \frac{N}{2}-\frac{N_\uparrow-N_\downarrow}{2}
-\lambda_{{\rm s}j} +j -\frac{1}{2}\right)
\right].
\label{dsip}
\end{eqnarray}
In the case of $\lambda_{{\rm s}2}=0$, where only one spinon is excited, we need to modify the above results as 
follows:
\begin{eqnarray}
\hspace{-1cm}
S^{zz}(Q,\omega)
&=&
\frac{1}{2^2N}
\frac{\Gamma[\frac{1}{2}] \Gamma[\lambda_{{\rm s}1}+1] 
\Gamma[\frac{N}{2}-\frac{N_\uparrow-N_\downarrow}{2}+1] 
\Gamma[\frac{N}{2}-\frac{N_\uparrow-N_\downarrow}{2}-\lambda_{{\rm s}1}+\frac{1}{2}]}
{\Gamma[\lambda_{{\rm s}1}+\frac{1}{2}] \Gamma[\frac{N}{2}-\frac{N_\uparrow-N_\downarrow}{2}+\frac{1}{2}] 
\Gamma[\frac{N}{2}-\frac{N_\uparrow-N_\downarrow}{2}-\lambda_{{\rm s}1}+1]} \nonumber \\
& & \times 
\delta(\omega-\Delta E_{\lambda}), 
\label{szzqwexception}
\end{eqnarray}
where $Q=2\pi \lambda_{{\rm s}1}/N$. The excitation energy $\Delta E_\lambda$ is given by $(\lambda_{{\rm s}1},\lambda_{{\rm s}2},\lambda_{\rm a})=(\lambda_{s1},0,2)$ in Eq.(\ref{dsip}).
We have checked the validity by comparison with numerical results up to $N=16$
~\cite{saiga00}
(see \ref{app.comparison}). In Figures \ref{comp.fig1} and \ref{comp.fig2}, we show
the results for $N=16$. From comparison with numerical results, the analytic expression 
of the two-spinon plus one-antispinon contribution can be applied  in the wider range of ($Q,\omega$) (see \ref{app.comparison}).  
The analytic expressions (\ref{eq:finiteszzexp1}) 
to (\ref{szzqwexception}) coincide with 
those for the Haldane-Shastry model\cite{arikawa00}.
{}From the above results for finite systems,
 we can derive the analytic expression of $S^{zz}(Q,\omega)$ in the thermodynamic limit (see Eq.(\ref{f:thermodynamiclimit})).
Note that
the contribution in the case of $\lambda_{{\rm s}2}=0$ 
vanishes in the thermodynamic limit. 
Thus we have proved analytically that in the momentum region $0< Q \le \min[\pi \bar{m},k_{{\rm F},\downarrow}]$, the structure factor 
$S^{zz}(Q,\omega)$ is not affected by hole doping. Here $k_{{\rm F},\sigma}$ is given by $\pi \bar{n}_\sigma$.
In this region, $S^{zz}(Q,\omega)$ diverges as $(\omega-\epsilon_{\rm s}(Q))^{-1/2}$, as the frequency approaches 
the lower edge corresponding to the spinon dispersion $\omega = \epsilon_{\rm s}(Q)$. 
The obtained $S^{zz}(Q,\omega)$ has the same form as the dynamical density-density correlation function of the spinless Sutherland model with coupling $\beta=2$ except the momentum range \cite{lesage95,ha95}. 
For $0< Q \le \min[\pi \bar{m},k_{{\rm F},\downarrow}]$,
the static 
structure factor $S^{zz}(Q)$ can be evaluated 
by the integration $\int d \omega S^{zz}(Q,\omega)$, which reproduces the expressions
presented in \ref{app.static}.

\section{Summary}
\label{sec.summary}
By use of the freezing technique on the Sutherland model with SU(2,1) supersymmetry and 
the replica type technique, we have obtained the dynamical spin structure factor $S^{zz}(Q,\omega)$ with $Q \le \min[\pi \bar{m}, k_{{\rm F},\downarrow}]$
in the supersymmetric {\itshape t-J} model with $1/r^2$ interaction.
The $S^{zz}(Q,\omega)$ has the same form as that of the Haldane-Shastry model in this small momentum region. 
In the thermodynamic limit, two spinons and one antispinon contribute to 
$S^{zz}(Q,\omega)$. 
Therefore,
$S^{zz}(Q,\omega)$ is not affected by hole doping in this region.
Thus we have proved the strong spin-charge separation in $S^{zz}(Q,\omega)$,
which was numerically obtained in the previous paper \cite{saiga00}.
{}From comparison with numerical results, we have found that the analytic expression of 
two-spinon plus one-antispinon contribution can be applied to the wider range of $(Q,\omega)$.  

\ack
The authors would like to thank Y. Kuramoto, Y. Kato, T. Yamamoto, R. Sasaki and P.J. Forrester for valuable discussions.
Y.S. is supported by
the 21st Century COE program ``ORIUM'' at Nagoya University from MEXT of Japan.

\appendix

\section{Dynamical charge structure factor $N(Q,\omega)$}
\label{app.nqw}
We derive the dynamical charge structure factor $N(Q,\omega)$ 
for $Q \le k_{F,\downarrow}$
\footnote{For the derivation of 
$N(Q,\omega)$, the condition $Q \le \pi \bar{m}$, i.e., 
Eq.(\ref{f:small1}) is not necessary.} 
by use of the replica type technique \cite{gangardt01}. 
The $N(Q,\omega)$ is defined by
\begin{eqnarray}
N(Q,\omega) & = & \sum_{\nu} |\langle \nu | n_{Q}| 0 \rangle |^2 
\delta(\omega-E_\nu+E_0),
\end{eqnarray} 
where $n_{Q} = \sum_l n_l e^{i Q l}/\sqrt{N}$.
As in our previous study\cite{arikawa99},  using the expansion coefficient $c_\lambda^{\rm h}$ in Eq.(\ref{f:expansioncoef1}), the dynamical charge structure factor
 $N(Q,\omega)$ can be expressed as 
\begin{eqnarray}
N(Q,\omega) & = & 
\frac{1}{N}
\sum_\lambda \left(c_\lambda^{\rm h}\right)^2
\frac{\langle K_\lambda, K_\lambda \rangle_N}{\langle K_{\lambda_{\rm GS}}, K_{\lambda_{\rm GS}} \rangle_N} \delta(\omega -\Delta E_\lambda).
\end{eqnarray}
for the momentum $Q>0$. Using the replica type technique \cite{gangardt01},
we can obtain the coefficient $c_\lambda^{\rm h}$.
By use of Eqs.(\ref{eq:replicares11}) and (\ref{separatedexpand}) with $(p,q)=(u,0)$,
the expansion coefficient  $c_\lambda^{\rm h}$ can be derived as
 \begin{eqnarray}
 c_\lambda^{\rm h}
 & = & 
 m \frac{\left[ 0 \right]^{\beta''}_{\nu^{\rm h}} \times f_{\nu^{\rm s}}(\bar{\beta}';0)}
{d'_{\nu^{\rm h}}(\beta'') d'_{\nu^{\rm s}}(\bar{\beta}') } 
= m \frac{\left[ 0 \right]^{\beta''}_{\nu^{\rm h}} }
{d'_{\nu^{\rm h}}(\beta'')} \delta_{(\nu^{\rm s}), (0^{N_\downarrow})},
\label{eq:expandhole}
 \end{eqnarray}
in the small momentum region $m \le ({N_{\downarrow}-1})/{2}$. 
Here we have used the property that $f_{\nu^{\rm s}}(\bar{\beta}';0)$
vanishes unless $\nu^{\rm s}=(0^{N_\downarrow})$.
In fact, as shown in Ref.\cite{arikawa99}, the excited states contributing to $N(Q,\omega)$
in this small momentum region are restricted to the case where $\lambda=\lambda_{\rm GS} +
(\nu^{\rm h},0^{N_\downarrow},0^{N_\uparrow})$. In this case,
the SU(2,1) 
Jack polynomial $K_{\lambda}(z,\beta)$ in Eq.(\ref{f:expansioncoef1}) has a form
$K_{\lambda}(z,\beta) = P_{\nu^{\rm h}}(z^{\rm h},\beta'') 
\times K_{\lambda_{\rm GS}}$. Therefore the
coefficient $ c_\lambda^{\rm h}$ can be derived
via Eq.(\ref{hanlonformula}) as well. 
Namely, the use of the replica type technique is not essential for the 
derivation
of $c^{\rm h}_{\lambda}$
in contrast to the case for $c^{\downarrow}_{\lambda}$.
The norm can be evaluated by the 
reduced formula Eq.(\ref{f:normsimple1}).
 In the strong coupling limit $\beta \rightarrow \infty$, the coupling parameter
 $\beta''$ becomes $1/2$.  In the thermodynamic limit, the dynamical 
 charge structure factor $N(Q,\omega)$ for
 $0 < Q \le k_{{\rm F},\downarrow}$ can be expressed as \cite{arikawa99}
 \begin{eqnarray}
N(Q,\omega) & = & \frac{Q^2}{\pi} 
\int_0^{2\pi-4k_F} d q_{\rm ah} \int_0^{k_{F,\downarrow}} d q_1
\int_0^{k_{F,\downarrow}} d q_2 \nonumber \\
& & \times \delta\left(Q-q_{\rm ah}-\sum_{j=1}^2 q_j\right) \delta\left(\omega-\epsilon_{\rm ah}(q_{\rm ah})-\sum_{j=1}^2 \epsilon_{\rm h}( q_j) \right)
\nonumber  \\
& & 
\times \frac{|q_1-q_2| \epsilon_{\rm ah}(q_{\rm ah})}{\prod_{j=1}^2 
\sqrt{\epsilon_{\rm h}(q_j)} 
\left(2 q_j+q_{\rm ah} \right)^2 }
\label{f:thermodynamiclimit2}
\end{eqnarray}
where  the Fermi momentum $k_{\rm F}$ is given by $k_{\rm F}=\pi \bar{n}/2$,
$\epsilon_{\rm h}(q)$ is the holon spectrum:
$\epsilon_{\rm h}(q) = q (v_{\rm c}+q)$ and 
$\epsilon_{\rm ah}(q)$ is the antiholon spectrum: $\epsilon_{\rm ah}(q) 
= q(v_{\rm c}-q/2)$.
Here the charge velocity $v_{\rm c}$ is $v_{\rm c}=\pi(1-\bar{n})$. 
This expression has the same form as
the dynamical density-density correlation function of the spinless Sutherland model with coupling parameter 
 $\beta=1/2$ . 
 
\section{Comparison with numerical results}
\label{app.comparison}

We make a comparison between analytic results and numerical ones in $S^{zz}(Q,\omega)$ \cite{saiga00}.
In Tables \ref{firsttable} and \ref{secondtable}, we present the cases
$(N,N_{\rm h},N_{\uparrow},N_{\downarrow})=(16,2,9,5)$ and  $(16,2,11,3)$,
respectively.
 Our analytic proof is restricted to the case where $Q \le 
 \min[k_{{\rm F},\downarrow},\pi \bar{m}]$.
 However, the analytical expression of  
 the two-spinon plus one-antispinon contribution can be applied in the wider range.
 As a result of hole doping, 
the integration ranges of the spinon momenta
in Eq.(\ref{f:thermodynamiclimit}) 
are changed to $0< q_i < k_{{\rm F},\downarrow}$ for $i=1$ and $2$.
 {}From comparison with numerical results\cite{saiga99,saiga00},
 we find that a similar fact occurs also in the $N(Q,\omega)$.
 Namely, although analytic derivation of $N(Q,\omega)$ is restricted to the region $0< Q < k_{{\rm F},\downarrow}$, the expression of  
 the (right-moving) two-holon plus one-antiholon contribution can be extended to the integration range shown in Eq.(\ref{f:thermodynamiclimit2}).

\begin{table}
\begin{center}
\caption{Comparison between analytic results and numerical ones \cite{saiga00} for $(N,N_{\rm h},N_{\uparrow},N_{\downarrow})=(16,2,9,5)$.\label{firsttable}}
\begin{tabular}{c c c c c c c}
\hline
$(\lambda_{s1},\lambda_{s2},\lambda_{a})$&
$Q/\pi$ & $\omega/t $ & \multicolumn{2}{c}{$I_\lambda$ (analytic)} & $I_\lambda$ (numeric) \\
\hline
(1,0,1)
&
$\frac{1}{8}$   &  $\frac{11\pi^2}{128}(\simeq 0.848169)$ &   
\multicolumn{2}{c}{$ \frac{3}{88} (\simeq 0.03409090)$ } 
&  0.03409080 \\
(2,0,1)
&
$\frac{1}{4}$   &  $\frac{9\pi^2}{64}(\simeq 1.38791)$ &   
\multicolumn{2}{c}{$ \frac{5}{99} (\simeq 0.0505050)$ } 
&  0.05050442 \\
(1,1,2)
&
   &  $\frac{3\pi^2}{16}(\simeq 1.85055)$ &   
\multicolumn{2}{c}{$ \frac{13}{528} (\simeq 0.02462121)$ } 
&  0.02462107 \\
(3,0,1)
&
 $\frac{3}{8}$  &  $\frac{21\pi^2}{128}(\simeq 1.61923)$ &   
\multicolumn{2}{c}{$ \frac{16}{231} (\simeq 0.06926406)$ } 
&  0.06926359 \\
(2,1,2)
&
   &  $\frac{31\pi^2}{128}(\simeq 2.39029)$ &   
\multicolumn{2}{c}{$ \frac{13}{352} (\simeq 0.036931818)$ } 
&  0.03693162 \\
(1,1,3)
&
   &  $\frac{39\pi^2}{128}(\simeq 3.00715)$ &   
\multicolumn{2}{c}{$ \frac{7}{352} (\simeq 0.01988636)$ } 
&  0.01988634 \\
&
$\frac{1}{2}$   & $\cdots$ &   
\multicolumn{2}{c}{$ \cdots $ } 
&   \\
(2,2,2)
&
   &  $\frac{5\pi^2}{16}(\simeq 3.08425)$ &   
\multicolumn{2}{c}{$ \frac{13}{6480} (\simeq 0.00200617)$ } 
&  0.00200648 \\
&
   & $\cdots$ &   
\multicolumn{2}{c}{$ \cdots $ } 
&   \\
\hline
\end{tabular}
\end{center}
\end{table}

%%%%%%%%%%%%%%%%%%%%%%%%%%%%%%%%%%%%

\begin{table}
\caption{Comparison between analytic results and numerical ones \cite{saiga00} for 
$(N,N_{\rm h},N_{\uparrow},N_{\downarrow})=(16,2,11,3)$.\label{secondtable}}
\begin{center}

\begin{tabular}{c c c c c c c}
\hline
$(\lambda_{s1},\lambda_{s2},\lambda_{a})$&
$Q/\pi$ & $\omega/t $ & \multicolumn{2}{c}{$I_\lambda$ (analytic)} & $I_\lambda$ (numeric) \\
\hline
(1,0,1)
&
$\frac{1}{8}$   &  $\frac{7\pi^2}{128}(\simeq 0.539744)$ &   
\multicolumn{2}{c}{$ \frac{1}{28} (\simeq 0.03571428)$ } 
&  0.035714172 \\
(2,0,1)
&
$\frac{1}{4}$   &  $\frac{5\pi^2}{64}(\simeq 0.771063)$ &   
\multicolumn{2}{c}{$ \frac{2}{35} (\simeq 0.057142857)$ } 
&  0.057142144 \\
(1,1,2)
&
   &  $\frac{\pi^2}{8}(\simeq 1.2337)$ &   
\multicolumn{2}{c}{$ \frac{3}{112} (\simeq 0.02678571)$ } 
&  0.026785570 \\
&
$\frac{3}{8}$   & $\cdots$ &   
\multicolumn{2}{c}{$ \cdots $ } 
&   \\
(1,1,3)
&
   &  $\frac{27\pi^2}{128}(\simeq 2.08187)$ &   
\multicolumn{2}{c}{$ \frac{5}{224} (\simeq 0.022321428)$ } 
&  0.022321422\\
&
$\frac{1}{2}$   & $\cdots$ &   
\multicolumn{2}{c}{$ \cdots $ } 
&   \\
(1,1,4)
&
   &  $\frac{5\pi^2}{16}(\simeq 3.08425)$ &   
\multicolumn{2}{c}{$ \frac{11}{560} (\simeq 0.019642857)$ } 
&  0.019642881\\
&
$\frac{5}{8}$   & $\cdots$ &   
\multicolumn{2}{c}{$ \cdots $ } 
&   \\
(1,1,5)
&
   &  $\frac{55\pi^2}{128}(\simeq 4.24085)$ &   
\multicolumn{2}{c}{$ \frac{1}{56} (\simeq 0.01785714)$ } 
&  0.017857178\\
&
   & $\cdots$ &   
\multicolumn{2}{c}{$ \cdots $ } 
&   \\
\hline
\end{tabular}
\end{center}
\end{table} 

\section{Static structure factors}
\label{app.static}
We consider the static structure factors $S^{zz}(Q)$ and $N(Q)$.
There are several ways to obtain these quantities.
If one knows the dynamical structure factors $S^{zz}(Q,\omega)$ and $N(Q,\omega)$, then the static structure factors can be obtained by integration
over $\omega$.
Gebhard and Vollhardt calculated the static structure factors
for $\bar{m}=0$, from the Gutzwiller wave function \cite{gebhard}.
For general $\bar{m}$, 
Forrester derived the analytic expressions of the equal-time two-point correlation functions \cite{peter,kuramotoprivate}
%%%%%%%%%%%%%%%%%%%%%%%%%%%%%%
\footnote{In Ref.\cite{peter}, the normalization factors are different.
When the spin correlation is divided into $C^{zz}(x)=C^{\rm hh}(x)/4+C^{\rm bh}(x)+
C^{\rm bb}(x)$, the correlation functions 
$C^{\rm hh}(x)$, $C^{\rm bh}(x)$ and $C^{\rm bb}(x)$ correspond to 
$\rho_{o}^2 h_{oo}(x)$, $\rho_{o}\rho_{\downarrow} h_{o\downarrow}(x)$ and $\rho_{\downarrow}^2 h_{\downarrow \downarrow}(x)$ in Ref.\cite{peter}, respectively. Tractable expressions for the correlation functions
were derived by Kuramoto\cite{kuramotoprivate}.}. 
%%%%%%%%%%%%%%%%%%%%%%%%%%%%%%
In the following, we obtain the the static structure factors
by 
Fourier transformation of these equal-time two-point correlation functions.

The equal-time two-point correlation functions are defined by $C^{zz}(x) \equiv \langle 0 | S^z_{x} S^z_{0} | 0\rangle$ and 
$C^{\rm hh}(x)\equiv \langle 0 | n_{x} n_{0} | 0\rangle$.
They are 
expressed as follows,
 \begin{eqnarray}
 C^{zz}(x) & = & \frac{\bar{m}^2}{4}+ \frac{\bar{n}-\bar{m}^2}{4} \delta_{x,0} \nonumber \\
 & &+ \frac{1-\delta_{x,0}}{4}\left[-[s_{\rm s}(x)]^2 +\left(\frac{d}{d x} s_{\rm s}(x) \right) 
 \int_0^{x} d u s_{-}(u) \right], \\
 C^{\rm hh}(x) & = &  \bar{n}^2 +\bar{n}(1-\bar{n})\delta_{x,0}   \nonumber \\
& & +(1-\delta_{x,0})\left[ -[s_{\rm c}(x)]^2 - \left(\frac{d}{d x} s_{\rm c}(x) \right) 
 \int_0^{x} d u s_{-}(u) \right], 
 \end{eqnarray}
where $s_{-}(x)$ is $s_{-}(x)=s_{\rm s}(x)-s_{\rm c}(x)$. 
$s_{\alpha}(x)$ ($\alpha =$c and s) are given by
\begin{eqnarray}
s_{\alpha}(x) & = & \frac{\sin v_{\alpha} x}{\pi x}.
\end{eqnarray}
By Fourier transformation we obtain the analytic expressions of $S^{zz}(Q)$ and $N(Q)$.
Taking into account of the Umklapp process, we obtain~\cite{kuramotoprivate}
 \begin{eqnarray*}
 S^{zz}(Q) & = & \frac{2v_{\rm s}-v_{\rm c}}{4\pi}
 + S_{I}(Q)+ S_{I}(2\pi-Q) + S_{II}(Q)+ S_{II}(2\pi-Q), \\
 N(Q) & = & \frac{v_{\rm c}}{\pi}
 + N_{I}(Q)+ N_{I}(2\pi-Q) + N_{II}(Q)+ N_{II}(2\pi-Q),
 \end{eqnarray*}
 for momentum $0<Q<2\pi$. 
The $ S_{I}(Q)$ is given by
 \begin{eqnarray}
 S_{I}(Q) & \equiv & 2\int_{0}^{\infty} d x \cos Q x \left[ 
 -\frac{1}{4}[s_{\rm s}(x)]^2 + \frac{1}{4}\left(\frac{d}{d x} s_{\rm s}(x) \right) 
 \int_0^{x} d u s_{\rm s}(u) 
 \right] \nonumber \\
 & = & \theta(2v_{\rm s}-Q) \left[ \frac{Q-2v_{\rm s}}{4\pi}-\frac{Q}{8\pi}\ln\left|1-\frac{Q}{v_{\rm s}}\right|\right],
 \end{eqnarray}
 where $\theta(x)$ is $\theta(x)=1$ for positive $x$, and $0$ otherwise. 
This contribution has the same form as the level-level correlation of the random matrices for 
symplectic ensembles\cite{sutherland4,mehta,efetov}.  
In fact, the $S^{zz}(Q)$ of the Haldane-Shastry model can be expressed by ${v_{\rm s}}/{(2\pi)}+
S_I(Q)+S_I(2\pi-Q)$ \cite{mucciolo}. 
The $S_{II}(Q)$ contributes for finite hole doping $(\bar{n}<1)$, 
which is given by
\begin{eqnarray}
S_{II}(Q) & \equiv & -\frac{1}{2} \int_{0}^{\infty} d x \cos Q x 
 \left(\frac{d}{d x} s_{\rm s}(x) \right) 
 \int_0^{x} d u s_{\rm c}(u) \nonumber \\
 & = & 
 \left\{
   \begin{array}{ll}
   \displaystyle\frac{v_{\rm c}}{4\pi}, & \mbox{for $0 < Q \le v_{\rm s}-v_{\rm c}$,} \\
   \displaystyle\frac{-Q+v_{\rm c}+v_{\rm s}}{8\pi} + \displaystyle\frac{Q}{8\pi} 
 \ln \left| \frac{Q-v_{\rm s}}{v_{\rm c}}\right|, & \mbox{for $v_{\rm s}-v_{\rm c} \le Q \le v_{\rm s}+v_{\rm c}$,} \quad \\
   0, & \mbox{for $Q \ge v_{\rm s}+v_{\rm c}$.}
   \end{array}
 \right.   
\end{eqnarray}
The divergence at $Q=\pi(1-\bar{m})$ in $S_{I}(Q)$ is removed by hole doping.
 The static spin structure factor has the same form  for the Haldane-Shastry model
 in the region $Q \le 2 k_{{\rm F},\downarrow}$.
 This region contains
 "$Q \le \min[\pi \bar{m}, k_{{\rm F},\downarrow}]$",
 where the $S^{zz}(Q,\omega)$ of the $1/r^2$ supersymmetric {\itshape t-J} model has the same form as that of the Haldane-Shastry model.
 For the momentum $0 < Q  \le \min[\pi \bar{m},k_{{\rm F},\downarrow}]$, 
 the $\omega$-integration of the $S^{zz}(Q,\omega)$  
 (see Eq.(\ref{f:thermodynamiclimit}))
 reproduces the above expression. 

 %%%%%%%%%%%%%%%%%%%%%%%%%%%%%
Next we consider the static charge structure factor $N(Q)$. The $N_I(Q)$ is given by
 \begin{eqnarray}
 N_{I}(Q) & \equiv & 2\int_{0}^{\infty} d x \cos Q x \left[ 
 -[s_{\rm c}(x)]^2 - \left(\frac{d}{d x} s_{\rm c}(x) \right) 
 \int_x^{\infty} d u s_{\rm c}(u) 
 \right] \nonumber \\
 & = &
 \left\{
   \begin{array}{ll}
   -\displaystyle\frac{v_{\rm c}}{\pi} +\displaystyle\frac{Q}{\pi} -\displaystyle\frac{Q}{2\pi}\ln \left| 1+ 
 \displaystyle\frac{Q}{v_{\rm c}}\right|, & \mbox{for $0< Q \le 2 v_{\rm c}$,} \\
   \displaystyle\frac{v_{\rm c}}{\pi} 
 -\displaystyle\frac{Q}{2\pi}\ln \left|  
 \displaystyle\frac{Q+v_{\rm c}}{Q-v_{\rm c}}\right|, & \mbox{for $2 v_{\rm c} \le Q$.}
   \end{array}
 \right.
 \end{eqnarray}
 This term has the same form as the level-level correlation of the random matrices for 
orthogonal ensembles\cite{mehta,efetov}. In fact, the static structure factor
of the Sutherland model with coupling parameter $\beta=1/2$ is given by $v_{\rm c}/\pi+N_I(Q)$
\cite{sutherland4}.  
The $N_{II}(Q)$ is given by
 \begin{eqnarray}
N_{II}(Q) & \equiv & 2 \int_{0}^{\infty} d x \cos Q x 
 \left(\frac{d}{d x} s_{\rm c}(x) \right) 
 \int_x^{\infty} d u s_{\rm s}(u) \nonumber \\
 & = &
 \left\{
   \begin{array}{ll}
   0, & \mbox{for $0 < Q \le v_{\rm s}-v_{\rm c}$,} \\
   \displaystyle\frac{Q}{2\pi}\ln\displaystyle\frac{Q+v_{\rm c}}{v_{\rm s}}-
 \displaystyle\frac{Q-v_{\rm s}+v_{\rm c}}{2\pi}, & \mbox{for $v_{\rm s}-v_{\rm c} \le Q \le v_{\rm s}+v_{\rm c}$,} \quad \\
   -\displaystyle\frac{v_{\rm c}}{\pi}+\displaystyle\frac{Q}{2\pi}\ln \displaystyle\frac{Q+v_{\rm c}}
{Q-v_{\rm c}}, & \mbox{for $Q \ge v_{\rm s}+v_{\rm c}$.}
   \end{array}
 \right.
\end{eqnarray}
For $0 < Q  \le k_{{\rm F},\downarrow}$, 
 the $\omega$-integration of  
the $N(Q,\omega)$ (see Eq.(\ref{f:thermodynamiclimit2}))
 reproduces the above expression. 
%%%%%%%%%%%%%%%%%%%%%%%%%%%%%%%%%%%%%%%%
\begin{figure}[ht]
%\begin{picture}(200,100)(-125,0)%
\begin{center}
\includegraphics[width=6cm]{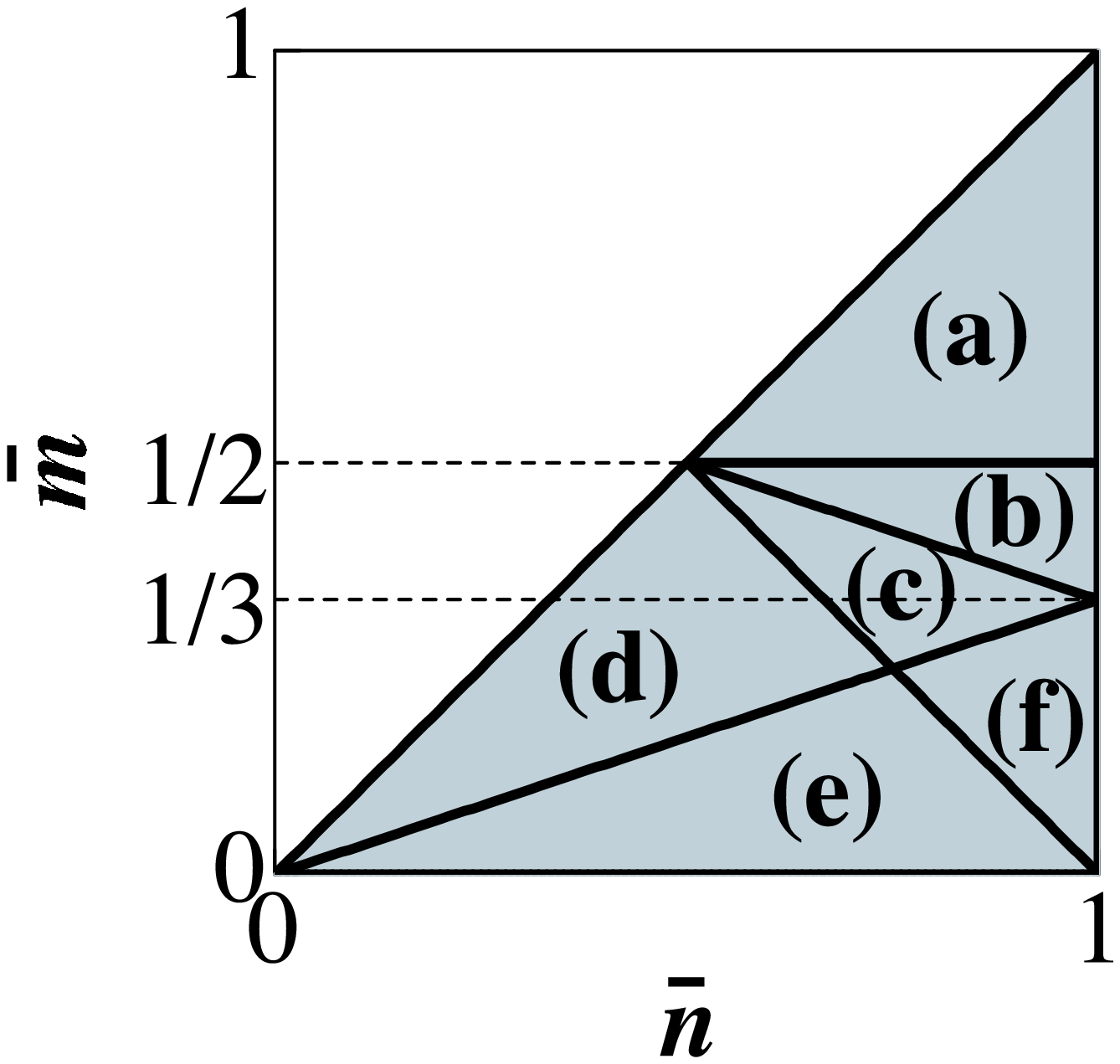}
\includegraphics[width=6cm]{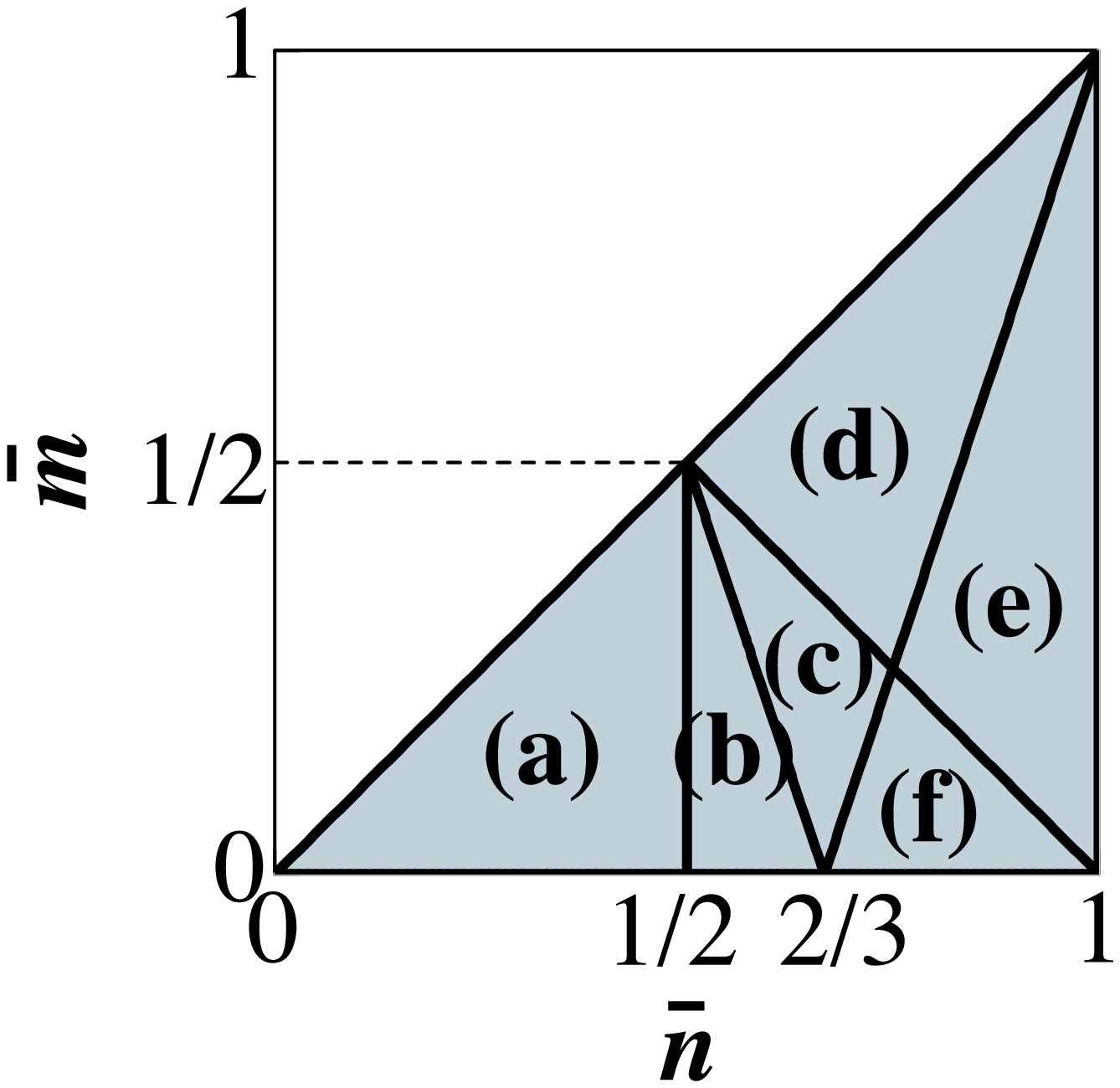}
\end{center}
%\end{picture}
\caption[]{
The shaded region shows the permissible one ($0 \le \bar{n} \le 1$ and $0 \le \bar{m} \le \bar{n}$) of the $(\bar{n},\bar{m})$ space in the {\it t-J} model.
This region can be classified into six regions with different expressions of $S^{zz}(Q)$ (left) and $N(Q)$ (right).
\label{fig:nqclassify}
}
\end{figure}
%%%%%%%%%%%%%%%%%%%%%%%%%%%%%%%%%%%%%%%%

%%%%%%%%%%%%%%%%%%%%%%
%%%.
We can rewrite $S^{zz}(Q)$ and $N(Q)$ more explicitly.
Because of the reflection symmetry against $Q=\pi$, it is enough to consider $Q \le \pi$.
For convenience, we define the following functions:
\begin{eqnarray}
S_{1}(x) & = & \frac{x}{4} -\frac{x}{8} 
\ln \left| \frac{1-\bar{m}-x}{1-\bar{m}}\right|, \nonumber \\
S_{2a}(x) & = & \frac{\bar{n}-\bar{m}}{8} + \frac{x}{8} - \frac{x}{8}
\ln  \frac{1-\bar{n}}{1-\bar{m}} , \nonumber \\
S_{2b}(x) & = & \frac{\bar{m}}{2} -\frac{x}{8} \ln \left| 
\frac{\bar{m}-1+x}{1+\bar{m}-x}\right|
-\frac{1}{4} \ln\left| \frac{1+\bar{m}-x}{1-\bar{m}}\right|, \nonumber \\
S_{3a}(x) & = & \frac{\bar{n}-1}{4} + \frac{x}{4}-\frac{x}{8}\ln\left| \frac{1-\bar{m}-x}{1-\bar{m}}\right|, \nonumber \\
S_{3b}(x) & = & \frac{\bar{n}+3\bar{m}}{8} -\frac{x}{8} 
+\frac{x}{8} \ln \left| \frac{1+\bar{m}-x}{1-\bar{n}}\right| 
-\frac{1}{4} \ln \left| \frac{1+\bar{m}-x}{1-\bar{m}}\right|, \nonumber \\
S_{4a}(x) & = & \frac{\bar{n}-2\bar{m}+1}{4}, \nonumber \\
S_{4b}(x) & = & \frac{\bar{n}+2\bar{m}-1}{4} -\frac{x}{8}
\ln \left| \frac{\bar{m}-1+x}{1+\bar{m}-x}\right|
-\frac{1}{4} \ln \left| \frac{1+\bar{m}-x}{1-\bar{m}}\right|, \nonumber \\
S_{4c}(x) & = & \frac{\bar{m}}{4} -\frac{1}{4} \ln \frac{1-\bar{n}}{1-\bar{m}}.
\end{eqnarray}
The expressions of $S^{zz}(Q)$ can classified into six cases (a) to (f) (see Figure \ref{fig:nqclassify} (left)).\\
(a) When $\bar{m} \ge 1/2$, we have 
\begin{equation}
S^{zz}(Q) =
\left\{
   \begin{array}{ll}
   S_{1}\left( Q/\pi  \right), & \mbox{for $0 <  Q \le \pi(\bar{n}-\bar{m})$,} \\
   S_{2a}\left( Q/\pi  \right), & \mbox{for $\pi(\bar{n}-\bar{m})
                                                    \le Q \le \pi(2-\bar{n}-\bar{m})$,} \\
   S_{3a}\left( Q/\pi  \right), & \mbox{for $\pi(2-\bar{n}-\bar{m})
                                                    \le Q \le 2 \pi (1-\bar{m})$,} 		\\												
   S_{4a}\left( Q/\pi  \right), & \mbox{for  $2 \pi (1-\bar{m})  \le Q  \le \pi$.} 															
   \end{array}
\right.   
\end{equation}
 (b) When $\bar{m} \le 1/2$ and $\bar{m} \ge -\bar{n}/3+2/3$, we have 
\begin{equation}
S^{zz}(Q) =
\left\{
   \begin{array}{ll}
   S_{1}\left( Q/\pi  \right), & \mbox{for $0 <  Q \le \pi(\bar{n}-\bar{m})$,} \\
   S_{2a}\left( Q/\pi  \right), & \mbox{for $\pi(\bar{n}-\bar{m})
                                                    \le Q \le \pi(2-\bar{n}-\bar{m})$,} \\
   S_{3a}\left( Q/\pi  \right), & \mbox{for $\pi(2-\bar{n}-\bar{m})
                                                    \le Q \le 2 \pi \bar{m}$,} 		\\
   S_{4b}\left( Q/\pi  \right), & \mbox{for  $2 \pi \bar{m}  \le Q  \le \pi$.} 															
   \end{array}
\right.   
\end{equation}
  (c) When $\bar{m} \le -\bar{n}/3+2/3$, $\bar{m} \ge -\bar{n}+1$ and $\bar{m} \ge \bar{n}/3$, we have 
\begin{equation}
S^{zz}(Q) =
\left\{
   \begin{array}{ll}
   S_{1}\left( Q/\pi  \right), & \mbox{for $0 <  Q \le \pi(\bar{n}-\bar{m})$,} \\
   S_{2a}\left( Q/\pi  \right), & \mbox{for $\pi(\bar{n}-\bar{m})
                                                    \le Q \le 2 \pi \bar{m}$,} \\
   S_{3b}\left( Q/\pi  \right), & \mbox{for $2 \pi \bar{m}
                                                    \le Q \le \pi(2-\bar{n}-\bar{m})$,} \\
   S_{4b}\left( Q/\pi  \right), & \mbox{for $\pi(2-\bar{n}-\bar{m}) \le Q  \le \pi$.} 															
   \end{array}
\right.   
\end{equation}
  (d) When $\bar{m} \le -\bar{n}+1$ and $\bar{m} \ge \bar{n}/3$, we have 
\begin{equation}
S^{zz}(Q) =
\left\{
   \begin{array}{ll}
   S_{1}\left( Q/\pi  \right), & \mbox{for $0 <  Q \le \pi(\bar{n}-\bar{m})$,} \\
   S_{2a}\left( Q/\pi  \right), & \mbox{for $\pi(\bar{n}-\bar{m})
                                                    \le Q \le 2 \pi \bar{m}$,} \\
   S_{3b}\left( Q/\pi  \right), & \mbox{for $2 \pi \bar{m}
                                                    \le Q \le \pi (\bar{n}+\bar{m})$,} 		\\
   S_{4c}\left( Q/\pi  \right), & \mbox{for  $\pi (\bar{n}+\bar{m}) \le Q  \le \pi$.} 															
   \end{array}
\right.   
\end{equation}
  (e) When $\bar{m} \le -\bar{n}+1$ and $\bar{m} \le \bar{n}/3$, we have 
\begin{equation}
S^{zz}(Q) =
\left\{
   \begin{array}{ll}
   S_{1}\left( Q/\pi  \right), & \mbox{for $0 <  Q \le 2 \pi \bar{m}$,} \\
   S_{2b}\left( Q/\pi  \right), & \mbox{for $2 \pi \bar{m}
                                                    \le Q \le \pi(\bar{n}-\bar{m})$,} \\
   S_{3b}\left( Q/\pi  \right), & \mbox{for $\pi(\bar{n}-\bar{m})
                                                    \le Q \le \pi (\bar{n}+\bar{m})$,} 		\\
   S_{4c}\left( Q/\pi  \right), & \mbox{for  $\pi (\bar{n}+\bar{m}) \le Q  \le \pi$.} 															
   \end{array}
\right.   
\end{equation}
 (f) When $\bar{m} \ge -\bar{n}+1$ and $\bar{m} \le \bar{n}/3$, we have 
\begin{equation}
S^{zz}(Q) =
\left\{
   \begin{array}{ll}
   S_{1}\left( Q/\pi  \right), & \mbox{for $0 <  Q \le 2 \pi \bar{m}$,} \\
   S_{2b}\left( Q/\pi  \right), & \mbox{for $2 \pi \bar{m}
                                                    \le Q \le \pi(\bar{n}-\bar{m})$,} \\
   S_{3b}\left( Q/\pi  \right), & \mbox{for $\pi(\bar{n}-\bar{m})
                                                    \le Q \le\pi (2-\bar{n}-\bar{m})$,} \\
   S_{4b}\left( Q/\pi  \right), & \mbox{for  $\pi (2-\bar{n}-\bar{m}) \le Q  \le \pi$.} 															
   \end{array}
\right.   
\end{equation}

%%%%%%%%%%%%%%%%%%%%%%%%%%%%%%
\noindent To describe $N(Q)$ as well, we define the following functions:
\begin{eqnarray}
N_{1}(x) & = & x - \frac{x}{2} \ln \left| \frac{1-\bar{n}+x}{1-\bar{n}} \right|, \nonumber \\
N_{2a}(x) & = & \frac{\bar{n}-\bar{m}}{2} + \frac{x}{2} 
+ \frac{x}{2}\ln  \frac{1-\bar{n}}{1-\bar{m}}, \nonumber \\
N_{2b}(x) & = & 2-2\bar{n}+\frac{x}{2}\ln\left| \frac{\bar{n}-1+x}
                                    {1-\bar{n}+x}\right|, \nonumber \\
N_{3a}(x) & = & -\bar{m} + x -\frac{x}{2} \ln \left| \frac{1+\bar{n}-x}{1-\bar{n}}\right| + \ln \left| \frac{1+\bar{n}-x}{1-\bar{m}}\right|, \nonumber \\
N_{3b}(x) & = & 2-\frac{3}{2}\bar{n}-\frac{\bar{m}}{2}-\frac{x}{2}+
                                   \frac{x}{2}\ln\left| \frac{\bar{n}-1+x}{1-\bar{m}}\right|,
								          \nonumber \\
N_{4a}(x) & = & 2\bar{n} -\bar{m}+ \ln \frac{1-\bar{n}}{1-\bar{m}}, \nonumber  \\
N_{4b}(x) & = & 2-2\bar{n} -\bar{m}+ 
                        \frac{x}{2} \ln \left|\frac{\bar{n}-1+x}{1+\bar{n}-x} \right| 
						+\ln \left|\frac{1+\bar{n}-x}{1-\bar{m}} \right|,  \nonumber \\
N_{4c}(x) & = & 1-\bar{n}.
\end{eqnarray}
%%%%%%%%%%%%%%%%%%%%%%%%%%%%%%%%%%%%%%%%
%%%.
The expressions of $N(Q)$ can be classified into six cases (a) to (f) (see Figure \ref{fig:nqclassify} (right)).\\
(a) When $\bar{n} \le 1/2$, we have 
\begin{equation}
N(Q) =
\left\{
   \begin{array}{ll}
   N_{1}\left( Q/\pi  \right), & \mbox{for $0 <  Q \le \pi(\bar{n}-\bar{m})$,} \\
   N_{2a}\left( Q/\pi  \right), & \mbox{for $\pi(\bar{n}-\bar{m})
                                                    \le Q \le \pi(\bar{n}+\bar{m})$,} \\
   N_{3a}\left( Q/\pi  \right), & \mbox{for $\pi(\bar{n}+\bar{m})
                                                    \le Q \le 2 \pi \bar{n}$,} 		\\
   N_{4a}\left( Q/\pi  \right), & \mbox{for  $2 \pi \bar{n}  \le Q  \le \pi$.} 															
   \end{array}
\right.   
\end{equation}
(b) When $\bar{n} \ge 1/2$ and $\bar{m} \le -3\bar{n}+2$, we have 
\begin{equation}
N(Q) =
\left\{
   \begin{array}{ll}
   N_{1}\left( Q/\pi  \right), & \mbox{for $0 <  Q \le \pi(\bar{n}-\bar{m})$,} \\
   N_{2a}\left( Q/\pi  \right), & \mbox{for $\pi(\bar{n}-\bar{m})
                                                    \le Q \le \pi(\bar{n}+\bar{m})$,} \\
   N_{3a}\left( Q/\pi  \right), & \mbox{for $\pi(\bar{n}+\bar{m})
                                                    \le Q \le 2 \pi (1-\bar{n})$,} 		\\
   N_{4b}\left( Q/\pi  \right), & \mbox{for  $2 \pi (1-\bar{n})  \le Q  \le \pi$.} 															
   \end{array}
\right.   
\end{equation}
(c) When $\bar{m} \ge 3\bar{n}-2$, $\bar{m}\le -\bar{n}+1$
and $\bar{m} \ge -3\bar{n}+2$, we have 
\begin{equation}
N(Q) =
\left\{
   \begin{array}{ll}
   N_{1}\left( Q/\pi  \right), & \mbox{for $0 <  Q \le \pi(\bar{n}-\bar{m})$,} \\
   N_{2a}\left( Q/\pi  \right), & \mbox{for $\pi(\bar{n}-\bar{m})
                                                    \le Q \le 2 \pi(1-\bar{n})$,} \\
   N_{3b}\left( Q/\pi  \right), & \mbox{for $2 \pi(1-\bar{n})
                                                    \le Q \le \pi (\bar{n}+\bar{m})$,} 		\\
   N_{4b}\left( Q/\pi  \right), & \mbox{for  $\pi (\bar{n}+\bar{m})  \le Q  \le \pi$.} 															
   \end{array}
\right.   
\end{equation}
(d) When $\bar{m}\ge -\bar{n}+1$ 
and $\bar{m} \ge  3\bar{n}-2$, we have 
\begin{equation}
N(Q) =
\left\{
   \begin{array}{ll}
   N_{1}\left( Q/\pi  \right), & \mbox{for $0 <  Q \le \pi(\bar{n}-\bar{m})$,} \\
   N_{2a}\left( Q/\pi  \right), & \mbox{for $\pi(\bar{n}-\bar{m})
                                                    \le Q \le 2 \pi(1-\bar{n})$,} \\
   N_{3b}\left( Q/\pi  \right), & \mbox{for $2 \pi(1-\bar{n})
                                                    \le Q \le \pi (2-\bar{n}-\bar{m})$,} 		\\
   N_{4c}\left( Q/\pi  \right), & \mbox{for  $\pi (2-\bar{n}-\bar{m})  \le Q  \le \pi$.} 															
   \end{array}
\right.   
\end{equation}
(e) When $\bar{m}\ge -\bar{n}+1$ 
and $\bar{m} \le  3\bar{n}-2$, we have 
\begin{equation}
N(Q) =
\left\{
   \begin{array}{ll}
   N_{1}\left( Q/\pi  \right), & \mbox{for $0 <  Q \le 2\pi(1-\bar{n})$,} \\
   N_{2b}\left( Q/\pi  \right), & \mbox{for $2\pi(1-\bar{n})
                                                    \le Q \le \pi(\bar{n}-\bar{m})$,} \\
   N_{3b}\left( Q/\pi  \right), & \mbox{for $\pi(\bar{n}-\bar{m})
                                                    \le Q \le \pi (2-\bar{n}-\bar{m})$,} 		\\
   N_{4c}\left( Q/\pi  \right), & \mbox{for  $\pi (2-\bar{n}-\bar{m})  \le Q  \le \pi$.} 															
   \end{array}
\right.   
\end{equation}
(f) When $\bar{m}\le -\bar{n}+1$ 
and $\bar{m} \le  3\bar{n}-2$, we have 
\begin{equation}
N(Q) =
\left\{
   \begin{array}{ll}
   N_{1}\left( Q/\pi  \right), & \mbox{for $0 <  Q \le 2\pi(1-\bar{n})$,} \\
   N_{2b}\left( Q/\pi  \right), & \mbox{for $2\pi(1-\bar{n})
                                                    \le Q \le \pi(\bar{n}-\bar{m})$,} \\
   N_{3b}\left( Q/\pi  \right), & \mbox{for $\pi(\bar{n}-\bar{m})
                                                    \le Q \le \pi (\bar{n}+\bar{m})$,} 		\\
   N_{4b}\left( Q/\pi  \right), & \mbox{for  $\pi (\bar{n}+\bar{m})  \le Q  \le \pi$.} 															
   \end{array}
\right.   
\end{equation}
In the limit $\bar{m} \rightarrow 0$, the above expressions of $S^{zz}(Q)$ and
$N(Q)$ reproduce the results in Ref.\cite{gebhard}.

%%%%%%%%%%%%%%%%%%%%%%%%%%%%%%%%%%%%%%%%%%

\end{document}